\documentclass[10pt]{article}

\usepackage{amsmath}
\usepackage{float}
\usepackage{graphicx}
\usepackage{color}
\usepackage{epsfig}  
\usepackage{graphics}  
\usepackage{times,mathptmx}
\usepackage{bm}

\newcommand{\rmd}{{\rm d}}
\newcommand{\rme}{{\rm e}}
\newcommand{\rmi}{{\rm i}}

\newcommand{\mbfo}{\mathbf{o}}
\newcommand{\mbfp}{\mathbf{p}}
\newcommand{\mbfr}{\mathbf{r}}
\newcommand{\mbfrho}{\mathbf{\rho}}
\newcommand{\mbfsigma}{\mathbf{\sigma}}
\newcommand{\elf}{{\cal E}_\perp}
\newcommand{\Elf}{{\cal E}}
\newcommand{\ELF}{{\mathbf{\cal E}}}
\newcommand{\mgf}{{\cal B}}
\newcommand{\MGF}{{\mathbf{\cal B}}}

\newcommand{\Ai}{\operatorname{Ai}\nolimits}
\newcommand{\Bi}{\operatorname{Bi}\nolimits}
\newcommand{\Ci}{\operatorname{Ci}\nolimits}

\newcommand{\He}{\operatorname{H}\nolimits}
\newcommand{\La}{\operatorname{L}\nolimits}
\newcommand{\sgn}{\operatorname{sgn}}

\newcommand{\bq}{\begin{equation}}
\newcommand{\eq}{\end{equation}}

\begin{document}

\title{New Mathematical Tools for Quantum Technology}
\author{
	C.~Bracher\\
	{\small Department of Physics and Astronomy, 
	California State University Long Beach}\\
	{\small 1250 Bellflower Blvd., Long Beach, CA 90840, USA}\\
	{\small E-mail address: cbracher@csulb.edu} \and  
	M.~Kleber  \\ 
	{\small Physik-Department T30,
	Technische Universit\"at M\"unchen}\\
	{\small James-Franck-Str., 85747 Garching, Germany} \\
	{\small E-mail address: mkleber@ph.tum.de} \and 
	T.~Kramer \\
	{\small Department of Physics,
	Harvard University}\\
	{\small 17 Oxford St., Cambridge, MA 02138, USA}\\
	{\small E-mail address: tobias.kramer@mytum.de}}

\maketitle

\begin{abstract}
Progress in manufacturing technology has allowed us to probe the behavior of devices on a smaller and faster scale than ever before.  With increasing miniaturization, quantum effects come to dominate the transport properties of these devices, between collisions, carriers undergo \emph{ballistic motion} under the influence of local electric and magnetic fields.  The often surprising properties of quantum ballistic transport are currently elucidated in ``clean'' atomic physics experiments.  From a theoretical viewpoint, the electron dynamics is governed by \emph{ballistic propagators} and \emph{Green functions}, intriguing quantities at the crossroads of classical and quantum mechanics.  Here, we briefly describe the propagator method, some ballistic Green functions, and their application in a diverse range of problems in atomic and solid state physics, such as photodetachment, atom lasers, scanning tunneling microscopy, and the quantum Hall effect.
\end{abstract}

\newpage
\section{Physics in small dimensions}
The laws of quantum mechanics provide a means for a successful interpretation of measurements and experiments on a microscopically small scale. It goes without saying that we can never ``observe'' directly what's going on in an atom or molecule. To understand what nature is telling us we must learn its language. Its grammar follows the mathematical rules of quantum mechanics. Without mathematical tools we would not be able to describe intriguing processes such as, for example the mapping of quantum states of light to intrinsic atomic states.  
Fortunately, the necessary formalism is often powerful, elegant and quite easy to comprehend. In our contribution we will demonstrate the revival of an established mathematical tool in the important field of quantum technology. This tool is known under the name of Green function or Green's function in honor of George Green.\footnote{Green was an almost entirely self-taught English mathematician and physicist who in 1828 published an essay entitled: ``{\it On the Applications of Mathematical Analysis to the Theories of Electricity and Magnetism.}'' In this essay he obtained integral representations for the solutions of problems connected with the Laplace operator.}

Many problems in electrodynamics, hydrodynamics, heat conduction, acoustics, etc., require the solutions of inhomogeneous linear differential equations. It is there where Green functions come to full power. The corresponding mathematical approach is the same in all branches of physics---as long as we are dealing with linear, ordinary or partial differential equations. In quantum mechanics, Green functions enjoy the advantage of having a physical meaning: The single-particle Green function $G(\mathbf{r}, \mathbf{r}\,';E)$ is the relative probability amplitude for a particle to move with energy $E$ from an arbitrary point $ \mathbf{r}\,' $ to another point $\mathbf{r}$.  Probability amplitudes are known to be essential in all kind of quantum problems. In his book on `{\it The Character of Physical Law}' Feynman \cite{Feynman1967a} notes that ``{\it \ldots everything that can be deduced from the ideas of the existence of quantum mechanical probability amplitudes, strange though they are, will work, \ldots one hundred percent.\ldots}''
\begin{figure}
\begin{center}
\includegraphics[width=0.6\textwidth]{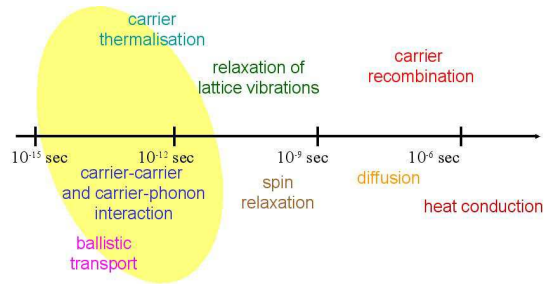} 
\caption{Timescales in semiconductors.
\label{fig_1}}
\end{center}
\end{figure}

In this tutorial we present some basic features of electron and atom motion in external fields. External electric, magnetic and even gravitational fields are well suited to control the motion of particles in quantum devices. It is not our purpose to dive into the technical depths of ultrasmall electronics research and technology. We only want to illustrate how useful {\it single-particle} Green functions and propagators can be for basic problems in quantum technology whenever a microscopic description of quantum transport is necessary. A single particle description is appropriate for devices with low particle densities where interaction processes can be neglected. But they are also useful in more general cases. Indeed, microscopically small particles travel freely on length scales of the order of the free mean path $\ell$, which is the distance that an electron travels before its initial phase is destroyed for whatever reasons \cite{Datta1997a,Imry2002a}. The mean free paths depend  strongly on the material under consideration and they are much affected by temperature. Particles that travel freely are called {\it ballistic} particles.

In Fig.~\ref{fig_1} we show timescales for the motion of electrons in a typical semiconductor. We invoke the uncertainty principle $ \Delta p \, \Delta x \sim \hbar $ to get a feeling for the time domain of ballistic motion. Motion with well defined momentum requires $ \Delta p \ll p$. For a quasi-classical description, the electrons are required to be well localized compared to the mean free path, i.~e., $ \Delta x \ll \ell = p\tau/m $ with  $ \tau $ being the time for ballistic motion (see Fig.1). It follows that $ \Delta p \, \Delta x \ll p\,\ell = p^2\tau/m $. Suppose we have thermal electrons; then we can use the equipartition law of classical statistics, $p^2\tau/m \sim k_{B}T\tau $ with $k_{B}$ being the Boltzmann constant and $T$ the temperature.  Comparing with the uncertainty principle we obtain $\tau \gg \hbar/(k_{B}T) $ which for room temperature is of the order $10^{-13}$ to $10^{-14}$ seconds. For thermal electrons this last inequality is frequently met. However, for electrons moving with high energies of the order eV the inequality is no longer fulfilled. In this case one has to treat ballistic transport fully quantum mechanically. In the following we present examples where quantum transport is essential. First, however, we review some useful mathematical tools in a nut shell.

Clearly, we cannot cite all relevant literature in this field. This would be an impossibly difficult and lengthy job. But the interested reader will find a wealth of literature in the research articles cited in this tutorial.

\section{Propagators and Green functions}

A time-dependent treatment of the flow of charge carriers is based on the time-dependent Schr\"odinger equation. In this context it is useful to summarize a few aspects of the initial-value problem for a wave function known at
$t=t_{0}$, 
\bq
\label{eq:bkk1}
\psi(\mathbf{r}, \, t=t_{0}) \,=\,\psi_{0}(\mathbf{r}) \,= \, \left\langle\mathbf{r} \vert \psi(t=t_{0})\right\rangle .
\eq
The corresponding ket vector $ |\psi(t)\rangle$ evolves according to the basic law of quantum mechanics, 
\bq
\label{eq:bkk2}
({\rm i}\hbar\partial_{t}-H) |\psi(t)\rangle = 0 .
\eq
The formal solution of (\ref{eq:bkk2}) is conveniently written in terms of the time evolution operator $U(t,t_{0})$: 
\bq
\label{eq:bkk3}
 |\psi(t)\rangle \, =\, U(t,t_{0}) |\psi(t_{0})\rangle.
\eq
In coordinate space (\ref{eq:bkk3}) reads
\bq
\label{eq:bkk4}
\langle\mathbf{r} |\psi(t)\rangle\,=\,\int {\rm d}^{3}\mathbf{r}'\,
\langle\mathbf{r} | U(t,t_{0}) |\mathbf{r}\,'\rangle
\langle\mathbf{r}\,' |\psi(t_{0})\rangle.
\eq
 The integral kernel of (\ref{eq:bkk4}) is called propagator $K$:
\label{eq:bkk5}
\bq
K(\mathbf{r},t |\mathbf{r}\,',t_{0})\,\equiv \,\langle\mathbf{r} | U(t,t_{0}) |\mathbf{r}\,'\rangle.
\eq
Obviously, $K(\mathbf{r},t |\mathbf{r}\,',t_{0})$ is the time evolution matrix $U(t,t_{0})$ in coordinate space representation.  From the last two equations, we have 
\bq
\label{eq:bkk6}
\lim_{t\rightarrow t_{0}}K(\mathbf{r},t | \mathbf{r}\,', t_{0})=
\delta^{3}(\mathbf{r} -\mathbf{r}\,')\,.
\eq  
The (time-) retarded  Green function must vanish for $t < t_0$. It is usually defined by
\bq
\label{eq:bkk7}
G(\mathbf{r},t;\mathbf{r}\,',t_{0})=\frac{1}{{\rm i}\hbar}\Theta(t-t_{0})
K(\mathbf{r},t |\mathbf{r}\,',t_{0}) \,.
\eq
From this definition, and the fact that the propagator is a solution of 
the time-dependent Schr\"odinger equation, the retarded Green function is seen to 
satisfy the differential equation 
\bq 
\label{eq:bkk8}
({\rm i}\hbar\partial_{t}-H)\,G(\mathbf{r},t;\mathbf{r}\,',t_{0})=\delta^{3}(\mathbf{r} -
\mathbf{r}\,') \delta(t-t_{0})
\eq
The delta function $\delta(t-t_{0})$ on the right hand side of (\ref{eq:bkk6}) originates 
from the step function $\Theta(t-t_{0})$ in the definition of $G$. 

For time-dependent Hamiltonians, the propagator will depend separately on $t$ and $t_0$. For time-independent Hamiltonians,  the propagator depends only on the time difference $t - t_0$.  In the latter case the Laplace transform of the propagator       
\bq
\label{eq:bkk9}
G(\mathbf{r},\mathbf{r}\,';E)\,=\,\frac{1}{{\rm i}\hbar}\int_{0}^{\infty} {\rm d}t\, \exp({\rm i}Et/\hbar)\, K(\mathbf{r},t  | \mathbf{r}\,',0)
\eq
generates the energy (-dependent) Green function,
\bq
\label{eq:bkk10}
G(\mathbf r,\mathbf r';E) = \lim_{\eta \rightarrow 0^+} \left\langle \mathbf r \left| \frac1{E-H+{\rm i}\eta} \right| \mathbf r' \right\rangle  .
\eq
$G(\mathbf{r},\mathbf{r}\,';E)$  is the amplitude for travel of a particle from $\mathbf{r}$ to $\mathbf{r}\,'$ out of a point source and, as a function of energy. This feature will emerge if we evaluate (\ref{eq:bkk10}) explicitly. We should also mention that the appearance of the infinitesimally small, positive imaginary term ${\rm i}\eta$ in (\ref{eq:bkk10}) has a simple reason: To enforce convergence of the integral in (\ref{eq:bkk9}) one has to replace $E$ by $E+{\rm i}\eta$. The physical meaning of such a small shift into the complex energy plane becomes evident if one evaluates (\ref{eq:bkk10}) for a free particle. The result \cite{Sakurai1994a},
\bq
\label{eq:bkk11}
G_{\rm free}(\mathbf{r},\mathbf{r}\,';E)\,=  -\frac{m}{2\pi \hbar^{2}}\; 
   \frac{\exp ({\rm i}k |\mathbf{r} -\mathbf{r}\,' |)}{ |\mathbf{r} -\mathbf{r}\,' |}, 
\eq
is well known from scattering theory: For $\mathbf{r}\,'$ fixed and $\mathbf{r}$ variable,
$G_{\rm free}$ describes an {\it outgoing} spherical wave that originates from a point source at $\mathbf{r}\,=  \mathbf{r}\,'$. Had we Fourier transformed the time-advanced Green function instead of the 
time-retarded Green function, we would, of course, have ended up with an
incoming spherical wave instead of an outgoing wave. 

Propagators contain all necessary information about the motion of a particle. 
Unfortunately it is not always possible to find a closed-form solution for $K$ or $G$. 
For potentials which are {\it at most quadratic in the coordinates},
the propagator assumes the canonical form \cite{Feynman1965a} 
\bq  
\label{K4}
K(\mathbf{r},t |\mathbf{r}\,',0)= A(t)\ 
               \exp\left[iS_{cl}(\mathbf{r},\mathbf{r}\,';t)/\hbar\right] , 
\eq
where $S_{cl}$ is the corresponding {\em classical action}, and where $A(t)$
is a time-dependent factor independent of the particle's position. 
However, nonquadratic potentials such as the Coulomb potential generally do not have the canonical form (\ref{K4}). Explicit expressions for propagators can be found, for example, in 
\cite{Feynman1965a,Schulman1981a,Kleinert1990a,Dodonov1992a,Nieto1992a,Kleber1994a,Grosche1998a}.\\

\begin{figure}[t]
\begin{center}
\hfil\includegraphics[width=0.3\textwidth]{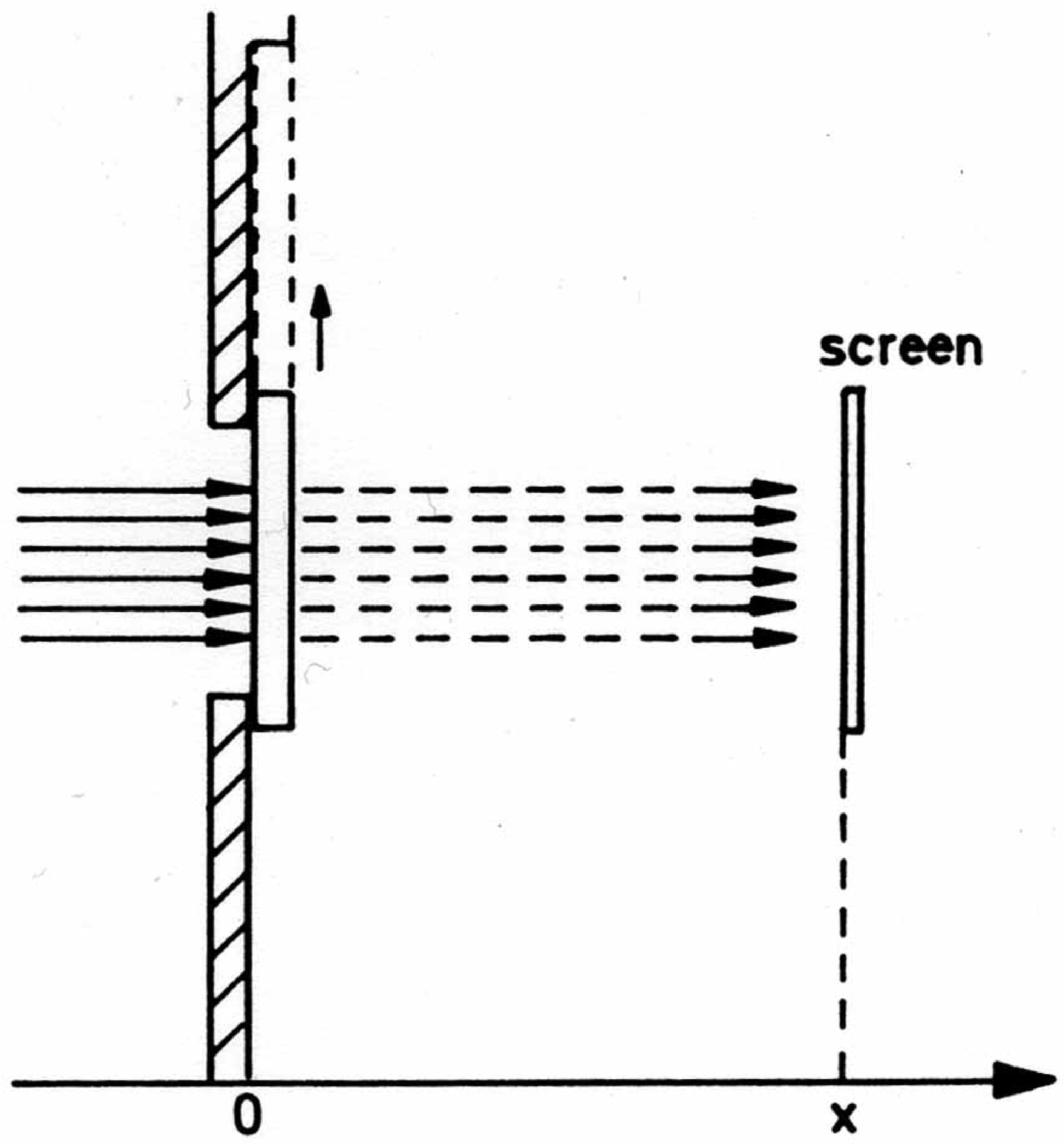}%
\hfil\includegraphics[width=0.6\textwidth]{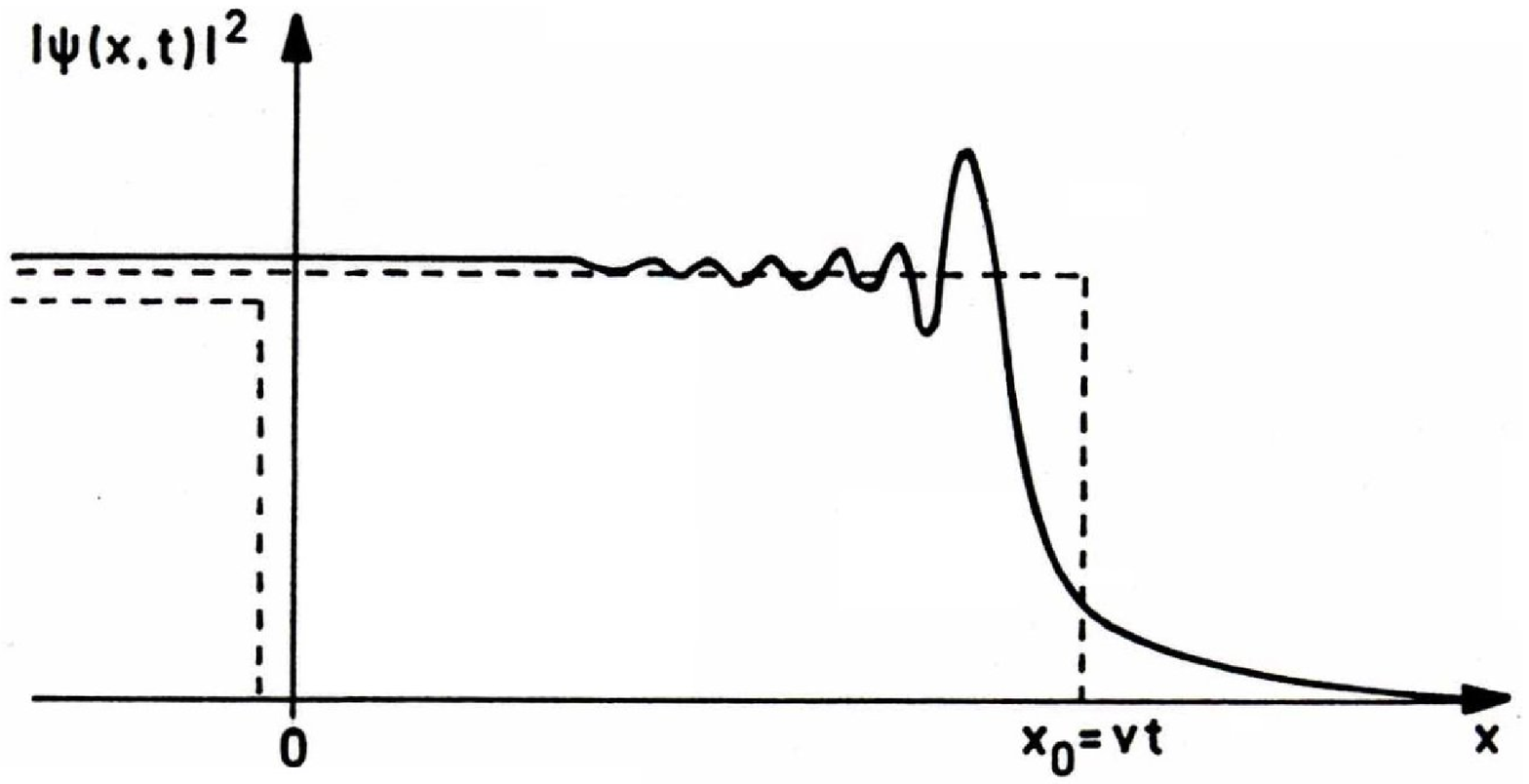}\hfil%
\caption{The Moshinsky shutter: The shutter is removed at $t_0 = 0$. The quantum particles start 
propagating towards the screen. When will the particles arrive at the screen? The distribution of the traveling particles is shown on the right hand side where the full line corresponds to the quantum result (17) and, where the classical propagation is represented by dotted lines.
\label{bkk_fig_2}}
\end{center}
\end{figure}

\paragraph{The Moshinsky shutter:} This example illustrates how the free propagator 
\bq
\label{bkk12}
K_{\rm free}(\mathbf{r},t | \mathbf{r}\,',t_{0}) = \left[\frac{m}{2\pi {\rm i}\hbar(t-t_{0})}\right]^{1/2}
\exp\left[\frac{{\rm i}m(\mathbf{r}-\mathbf{r}\,')^{2}}{2\hbar(t-t_{0})}\right] ,
\eq
is used to solve an initial value problem that describes the flow of quantum particles. It is of interest in the context of the quantum mechanical propagation of a signal.  
Moshinsky \cite{Moshinsky1952a} has analyzed the spreading of such a signal.  
He considered a monochromatic beam of noninteracting particles of mass $m$ and
energy $E_{k}$. The particles are supposed to move parallel to the $x$--axis 
from left to right. The beam is stopped (and absorbed) by a shutter at $x=0$ (see Fig.~\ref{bkk_fig_2}). 
The signal is given at $t=t_{0}=0$ when the shutter is opened. 

The sudden removal of the shutter marks the beginning of a ``quantum race'' where the particles run along the positive $x$--axis. In order to elucidate the spreading of the signal all one has to do is to 
calculate $\psi(x,t)$ starting with 
\bq
\psi_{0}(x) = \psi(x,t=0) = \Theta(-x) {\rm e}^{{\rm i}kx}.
\eq
Using (\ref{eq:bkk4}) one finds 
\bq
\label{eq:bkk13}
\langle x |\psi(t\geq 0)\rangle \,=\,\operatorname{M}\left(x; k; \hbar t/m \right),
\eq
where the Moshinsky function M is defined in 
terms of the complementary error function \cite{Abramowitz1965a},
\bq
\operatorname{M}(x;k;\tau)=\frac{1}{2} \exp\bigl( {\rm i}kx - {\rm i}k^{2}\tau/2 \bigr)\; 
\operatorname{erfc}\left[\frac{x-k\tau}{(2{\rm i}\tau)^{1/2} }\right]  
\eq
with ${\rm i}^{1/2} =\exp({\rm i}\pi/4)$.
  
An interesting property of (\ref{eq:bkk13}) is revealed when we evaluate the particle number probability. Introducing  
$u=(\hbar kt/m - x)/(\pi\hbar t/m)^{1/2}$, we obtain 
\bq
\label{eq:14}
\bigl| \langle x |\psi(t)\rangle \bigr|^{2}=\frac{1}{2}\left\{\left[\frac{1}{2}+
\mbox{C}(u)\right]^{2}+ \left[\frac{1}{2}+\mbox{S}(u)\right]^{2} \right\}.
\eq
The functions $\operatorname{C}(u)$ and $\operatorname{S}(u)$ are the well-known Fresnel integrals \cite{Abramowitz1965a}.  The corresponding probability pattern is called diffraction in time because it arises when the shutter is opened for a finite time $t$. Although transient effects are important by themselves \cite{DelCampo2005a} we won't discuss them in more detail here. In what follows, we will discuss stationary quantum transport. 

\section{Quantum sources}
\label{sec:Multi}

In real-space representation, propagators and Green functions describe the motion of quantum particles from some initial point $\mathbf{r}\,'$ to a final point $\mathbf{r}$. But where do the particles come from? One may think of two different situations: i) the particles have been around all the time like electrons in an atom, or, ii) the particles are generated by a source, a situation which is quite familiar from scattering theory where a beam of particles is generated by an accelerator in a region far away from the target. In mesoscopic physics and nanotechnology, however, there is usually no such large spatial separation. Let us motivate the introduction of coherent quantum sources of particles and illustrate their properties by means of an example.

\subsection{Photoelectrons emitted from a quantum source}
\label{sec:Multi1}

\begin{figure}[t]
\centerline{\hfil\includegraphics[width=0.4\textwidth]{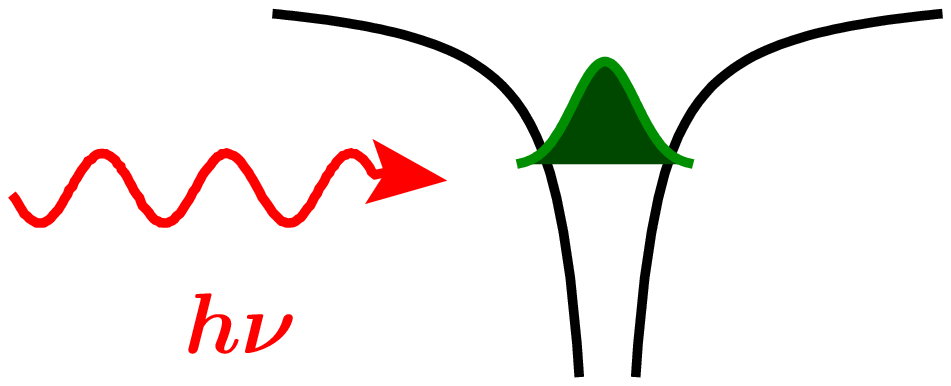}
\hfil\includegraphics[width=0.4\textwidth]{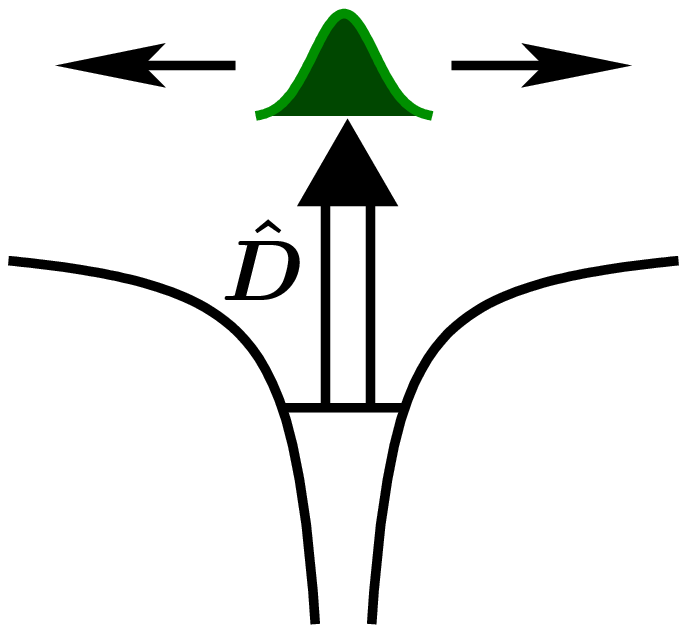}\hfil}
\caption{Two steps to create a photoelectron: Step 1 (left panel): The photon 
transfers its energy to the initially bound electron. 
Step 2 (right panel): The photoelectron escapes from the absorption region.
\label{bkk_fig_3}} 
\end{figure}

We may consider the photoeffect as a two-step process as illustrated in Fig.~\ref{bkk_fig_3}.  The time evolution of the emitted electron is of course governed by the rules of quantum mechanics. In the absence of any interaction between photon and electron, the electron under consideration is attached to the atom and is described by the bound-state wave function $\psi_{\rm atom}(\mathbf{r})$. 
Let us consider a dilute gas of independent atoms where the interaction of the photoelectron with neighboring atoms can be neglected.  In the presence of a photon field this wave function will obtain a small scattering component $\psi_{\rm sc}(\mathbf{r})$ that allows the electron to leave the atom. For a dipole-allowed transition, the dipole operator $\hat{D}(\mathbf{r})\, \propto \hat{\mathbf{\epsilon}}\cdot\mathbf{r}$ is responsible for transferring the electron from its initial bound state $|\psi_{\rm atom}\rangle$  to a continuum state $|\psi_{\rm sc}\rangle$ . Under steady-state conditions with many atoms (each having the Hamiltonian $\hat{H}_{\rm atom}$) and weak monochromatic light we must solve the problem:
\bq
\label{eq:bkk14}
\Bigl[E-\hat{H}_{\rm atom} -  \hat{H}_{\rm rad} - \hat{D}\cdot (\hat{a} + \hat{a}^{\dagger})\Bigr]\Bigl( |\psi_{\rm atom}\rangle \, |1\rangle +  |\psi_{\rm sc}\rangle \, |0 \rangle \Bigr) = 0 
\eq 
where $E = E_{\rm atom} + h\nu $ is the energy sum of electron and photon. The unperturbed Hamilton operator of the radiation field with field operators $\hat{a}$ and $\hat{a}^{\dagger}$ is denoted by $\hat{H}_{\rm rad}$, with the zero-point energy being subtracted. As usual, $|1\rangle$ characterizes the presence of the photon  and $|0\rangle$ its absence after absorption. Projection onto the zero-photon state $\langle 0 |$ yields the desired equation for the scattering solution,  
\bq 
\label{eq:bkk16}
\Bigl[ E-\hat{H}_{\rm atom}(\mathbf{r})\Bigr]\psi_{\rm sc}(\mathbf{r}) =  \, \hat{D}(\mathbf{r}) \psi_{\rm atom}(\mathbf{r}) \equiv  \sigma(\mathbf{r}).  
\eq
We can interpret $\hat{D}|\psi\rangle$ as a {\it source function} $|\sigma\rangle$ for the photoelectrons: The dipole operator {\it prepares} the electron in a continuum state but can be neglected once the electron has left the atom.

\subsection{Currents generated by quantum sources}
\label{sec:Multi2}

The last two equations can be generalized to a situation where the scattered particle experiences some final-state interaction. For example, the presence of a final-state Coulomb interaction or of an external field can be readily taken into account in Eq.~(\ref{eq:bkk16}) by writing  
\bq
\label{eq:Multi1.2}
\Bigl[ E - H_0 - W(\mathbf r) \Bigr] \psi_{\rm sc}(\mathbf r) = \sigma(\mathbf r)  \;,
\eq
where $H_0$ is the Hamiltonian of a free particle and where $W(\mathbf r)$ represents the interaction 
of the emitted particle (for example the photoelectron) with its environment. 
Here and in the following we omit the hat symbol for the operator $(H_0 \equiv \hat{H}_0)$. In analogy to other inhomogeneous field equations, e.~g., Maxwell's equations, the right-hand term $\sigma(\mathbf r)$ in (\ref{eq:Multi1.2}) is again identified as a source for the scattered wave $\psi_{\rm sc}(\mathbf r)$.

We now turn to the mathematical aspects of (\ref{eq:Multi1.2}).  Introducing the energy Green function $G(\mathbf r,\mathbf r';E)$ for the Hamiltonian $H$ defined via \cite{Economou1983a}:
\begin{equation}
\label{eq:Multi1.3}
\left[ E - H_0 - W(\mathbf r) \right] G(\mathbf r,\mathbf r';E) = \delta^3(\mathbf  r - \mathbf r') \;,
\end{equation}
a solution to (\ref{eq:Multi1.2}) in terms of a convolution integral reads:
\begin{equation}
\label{eq:Multi1.4}
\psi_{\rm sc}(\mathbf r) = \int {\rm d}^3r'\, G(\mathbf r,\mathbf r';E) \sigma(\mathbf r') \;.
\end{equation}
In general, this result is not unique. However, any two solutions $\psi_{\rm sc}^{1}(\mathbf r)$ and $\psi_{\rm sc}^{2}(\mathbf r)$ differ only by an eigenfunction $\psi_{\rm hom}(\mathbf r)$ of the homogeneous Schr\"odinger equation, with $H = H_0 + W$ and  $H \psi_{\rm hom}(\mathbf r) = E \psi_{\rm hom}(\mathbf r)$.  The ambiguity in $\psi_{\rm sc}(\mathbf r)$ is resolved by the demand that $G(\mathbf r,\mathbf r';E)$ presents a retarded solution characterized by outgoing-wave behavior as $r\rightarrow\infty$.  Formally, this enforces the same choice as in Eq.~(\ref{eq:bkk10}). It is then easy to decompose the Green function (\ref{eq:bkk10}) into real and imaginary parts
\begin{equation}
\label{eq:Multi1.5}
G(\mathbf r,\mathbf r';E) = 
\left\langle \mathbf r \left| \operatorname{PP} \left( \frac1{E-H} \right) - {\rm i}\pi\delta(E-H) \right| \mathbf r' \right\rangle \;,
\end{equation}
where $\mbox{PP}(\ldots)$ denotes the Cauchy principal value of the energy integration.

Defining the current density in the scattered wave in the usual fashion by $\mathbf j(\mathbf r) = \hbar \Im[\psi_{\rm sc}(\mathbf r)^* \mathbf{\nabla} \psi_{\rm sc}(\mathbf r)]/M$ (where for simplicity we omitted the vector potential $\mathbf A(\mathbf r)$, see \cite{Kramer2001a}), the inhomogeneous Schr\"odinger equation (\ref{eq:Multi1.2}) gives rise to a modified equation of continuity \cite{Bracher1997a,Kramer2002a}:
\begin{equation}
\label{eq:Multi1.6}
\mathbf{\nabla}\cdot\mathbf j(\mathbf r) = - \frac2\hbar \Im\left[ \sigma(\mathbf r)^* \psi_{\rm sc}(\mathbf r) \right] \;,
\end{equation}
where $\Im\left[ x \right]$ stands for the imaginary part of $x$. Thus, the inhomogeneity $\sigma(\mathbf r)$  acts as a source for the particle current $\mathbf j(\mathbf r)$.  By integration over the source volume, and inserting (\ref{eq:Multi1.4}), we obtain a bilinear expression for the total particle current $J(E)$, i.~e., the total scattering rate:
\begin{equation}
\label{eq:Multi1.7}
J(E) = - \frac2\hbar \Im\left[ \int {\rm d}^3r \int {\rm d}^3r' \sigma(\mathbf r)^* G(\mathbf r,\mathbf r';E) \sigma(\mathbf r') \right] \;.
\end{equation}
Some important identities concerning the total current $J(E)$ are most easily recognized in a formal Dirac bra-ket representation.  In view of (\ref{eq:Multi1.5}), we may express $J(E)$ by
\begin{equation}
\label{eq:Multi1.8}
J(E) = -\frac2\hbar \Im\left[ \left\langle \sigma \left| G \right| \sigma \right\rangle \right] = \frac{2\pi}\hbar \left\langle \sigma \left| \delta(E-H) \right| \sigma \right\rangle \;,
\end{equation}
from which the sum rule immediately follows \cite{Kramer2002a}:
\begin{equation}
\label{eq:Multi1.9}
\int_{-\infty}^{\infty} {\rm d}E\, J(E) = \frac{2\pi}\hbar \langle\sigma|\sigma\rangle = \frac{2\pi}\hbar \int {\rm d}^3r |\sigma(\mathbf r)|^2 \;,
\end{equation}
(provided this integral exists).

\subsection{Recovering Fermi's golden rule} 
\label{sec:Multi3}

In order to connect Eq.~(\ref{eq:Multi1.7}) to the findings of conventional scattering theory, we display $J(E)$ in an entirely different,  yet wholly equivalent fashion.  Employing a complete orthonormal set of eigenfunctions $| \psi_{\rm fi} \rangle$ of the Hamiltonian $H$, $\delta(E-H)| \psi_{\rm fi} \rangle = \delta(E-E_{\rm fi})| \psi_{\rm fi} \rangle$ follows, and replacing $| \sigma \rangle = D(\mathbf r)| \psi_{\rm atom} \rangle$ (\ref{eq:Multi1.5}), we may formally decompose (\ref{eq:Multi1.8}) into a sum over eigenfunctions:
\begin{equation}
\label{eq:Multi1.10}
J(E) = \frac{2\pi}\hbar \sum_{\rm fi} \delta(E-E_{\rm fi}) \left| \left\langle \psi_{\rm fi} |D(\mathbf r)| \psi_{\rm atom} \right\rangle \right|^2 \;.
\end{equation}
Thus, Fermi's golden rule is recovered.  Another noteworthy consequence of (\ref{eq:Multi1.7}) and (\ref{eq:Multi1.8}) emerges in the limit of pointlike sources, $\sigma(\mathbf r) \sim C \delta(\mathbf r-\mathbf R)$.  We then find \cite{Bracher1997a}
\begin{equation}
\label{eq:Multi1.11}
J(E) = -\frac2\hbar |C|^2 \Im[G(\mathbf R,\mathbf R;E)] = \frac{2\pi}\hbar |C|^2 n(\mathbf R;E) \;,
\end{equation}
where $n(\mathbf R;E) = \sum_{\rm fi} \delta(E-E_{\rm fi}) |\psi_{\rm fi}(\mathbf R)|^2$ is the local density of states of $H$ at the source position $\mathbf R$.  Equation (\ref{eq:Multi1.11}) forms the theoretical basis of the Tersoff--Hamann description of scanning tunneling microscopy \cite{Bracher1997a,Tersoff1983a}. The advantage of the formulation in terms of quantum sources over the traditional Fermi's golden rule approach (which involves an integral over the final states) is that it emphasizes the dynamical aspects of the propagation in real space and opens the possibility to a semiclassical calculation of photocurrents with closed-orbit theories \cite{Kleppner2001a,Du1988a}.

\subsection{Photodetachment and Wigner's threshold laws}
\label{sec:Multi4}

To find out how we can use the formalism for real physics we continue our discussion of the photoeffect. Applying the photoelectric effect to negative ions means that the emitted electron only weakly interacts with the remaining neutral atom. 
\begin{figure}
\begin{center}
\hfil%
\includegraphics[height=0.4\textwidth]{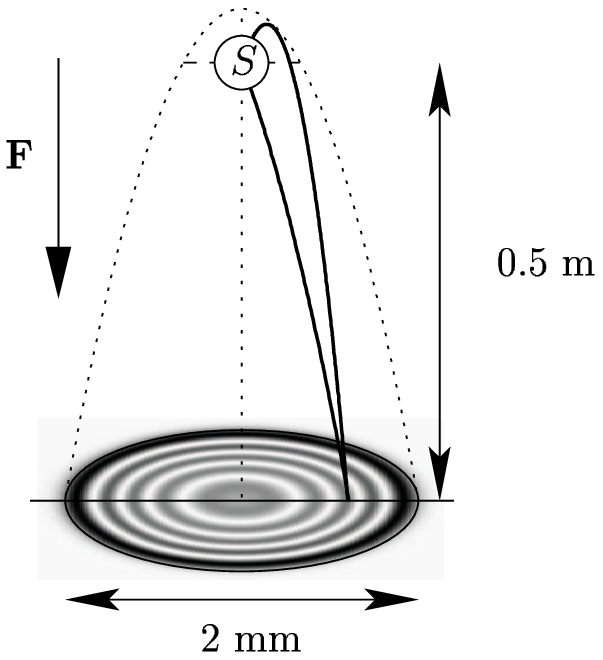}%
\hfil%
\includegraphics[height=0.4\textwidth]{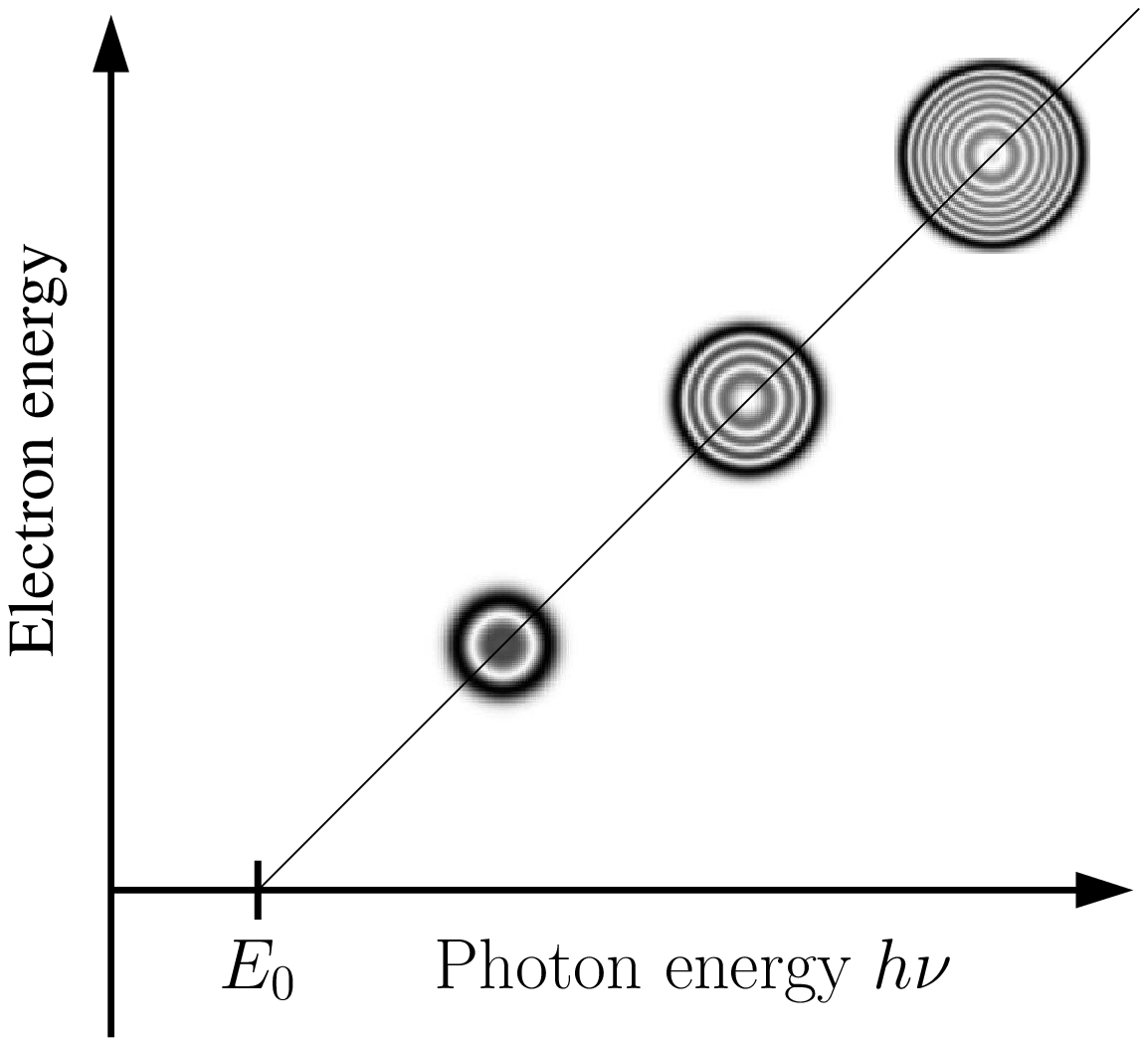}%
\hfil
\end{center}
\caption{\label{fig:blondel}
Near-threshold detachment of oxygen ions: ${\rm O}^- \rightarrow {\rm O}+e^-$ in the presence of a homogeneous electric force field $F = e\cal{E}$. The two possible classical trajectories for a photoelectron leading from the source (marked by $S$) to any destination will give rise to interference on a distant detector screen. The fringe pattern in the current distribution depends sensitively on the energy. By counting the number of fringes the binding energy $E_0$ of the outer electron can be determined from Einstein's law \cite{Blondel1996a,Blondel1999a,Bracher2003a}.}
\end{figure}
Just as in Young's double-slit experiment, the fringe pattern in the current profile can be interpreted 
as interference between the two classical trajectories, here of a particle in a constant force 
field \cite{Bracher2003a,Demkov1982a}. From the interference pattern one can determine the kinetic energy 
of the electrons and plot it against the photon energy to check Einstein's law (right panel 
of Fig.~\ref{fig:blondel}). Near threshold, the photoelectron has very little kinetic energy and, hence, 
a large de Broglie wavelength. In the absence of a final-state Coulomb interaction the relevant Green 
function is that of a particle falling freely in a constant field \cite{Bracher2003a,Bracher1998a}.
Experimental results are reported in Ref.~\cite{Blondel1999a}, Fig.~4. They show a highly accurate 
verification of Einstein's law which can be used to obtain the binding energy of O$^-$ with unprecedented 
accuracy.´

As noted above, we can interpret $D(\mathbf r)|\psi_{\rm atom}\rangle$ as a source $|\sigma\rangle$ for the photoelectrons. Expanding the source in terms of multipoles $\sigma_{lm}(r)$ and, taking into account that for O$^-$ the photoelectron leaves the atom near threshold in an $s$--wave continuum state, we retain only the $l=0$ component of the source by writing  $\sigma(\mathbf{r}) = \sigma_{00}(r)/\sqrt{4\pi}$ . 
In the absence of external fields, the free Green function (\ref{eq:bkk11}), a outgoing spherical wave, yields after multipole expansion
\begin{equation}
J(E)\propto k{\left[\int_0^\infty\frac{r\,\rmd r}{k}\sin(k r) \sigma_{00}(r)\right]}^2.
\end{equation}
For $E=\hbar^2k^2/(2m)\rightarrow 0$ it follows that $J(E)\propto k$. This result is independent of the form of the atomic source and it reflects Wigner's threshold law \cite{Bracher2003a,Wigner1948a}. It provides an alternative way to determine the electron affinity of a negative ion.

A similar analysis applies to the more complicated Green function in an electric force field $F = e \cal{E}$, one of the few quantum problems in more than one dimension that have exact solutions:
\begin{equation}
\label{GreenField}
G(\mbfr,\mbfo;E)=\frac{m}{2\hbar^2 r}
\left[ \Ci(\alpha_+)\Ai'(\alpha_-) - \Ci'(\alpha_+)\Ai(\alpha_-) \right].
\end{equation}
Here, $\beta^3 = m/(2\hbar F)^2$, $\alpha_\pm = -\beta[2E + F(z \pm r)]$, and $\Ai(u)$ and $\Ci(u) = \Bi(u) + {\rm i}\Ai(u)$ denote Airy functions \cite{Abramowitz1965a}.
It leads to a modified Wigner law for the $s$--wave absorption cross section near threshold \cite{Kramer2002a}:
\begin{equation}
J(E) \propto \Ai'(-2\beta E)^2 + 2\beta E \Ai(-2\beta E)^2.
\end{equation}
A static electric field opens up a sub-threshold ($E<0$) tunneling regime that has been confirmed by experiment \cite{Gibson2001a}.

Emission of particles from pointlike sources has been considered in the literature \cite{Rodberg1967a} long before the advent of mesoscopic physics. It was Schwinger \cite{Schwinger1973a} who introduced sources as a means of describing quantum dynamics in the context of emission and absorption of light.

\section{Spatially extended sources: The atom laser}
\label{sec:GaussianSource}

Atomic electron sources are usually sufficiently small to be considered pointlike.  A different situation arises when particles are coherently emitted from an extended region in space.  An example for such a ``fuzzy'' source is the continuous \emph{atom laser}, a beam of ultracold atoms fed by a Bose--Einstein condensate (BEC) \cite{Bloch1999a}.  In the experiment, only atoms in a specific Zeeman substate ($m=-1$) are magnetically trapped and form a BEC.  Application of a suitably tuned radiofrequency (RF) field will cause transitions into another magnetic substate of the atoms ($m=0$) that is not subject to the trapping potential.  Under the influence of gravity, these ``outcoupled'' atoms fall freely from the trap region and form a coherent, continuous atom laser ``beam.''  In our language, the macroscopic BEC wave function $\psi_0(\mathbf r)$ serves as the source $\sigma(\mathbf r)$ and corresponds to the atomic bound state $\psi_{\rm atom}(\mathbf r)$ in Eq.~(\ref{eq:bkk14}), whereas the outcoupled beam $\psi(\mathbf r)$ of accelerating atoms takes over the role of the scattered wave $\psi_{\rm sc}(\mathbf r)$, akin to photodetachment in an electric field (Sec.~\ref{sec:Multi4}).  From a theoretical viewpoint, the only essential difference is the macroscopic size of the source.
\begin{figure}[t]
\centerline{\includegraphics[width=0.9\textwidth]{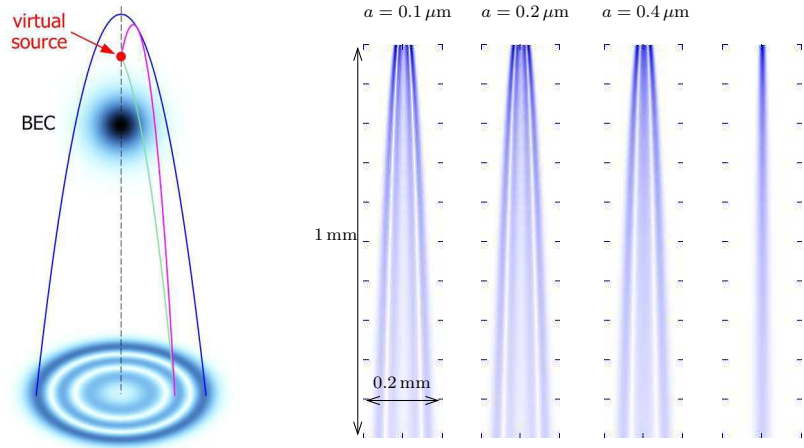}}
\caption{Left panel: A Gaussian source of freely falling particles can be replaced by a \emph{virtual} point source of particles with the same energy, located upstream from the actual extended source. Right panels:
Size dependence of the beam profile for Gaussian BEC sources with different widths $a$.
\label{fig:virtual+beam}}
\end{figure}
\begin{figure}[t]
\includegraphics[width=0.47\textwidth]{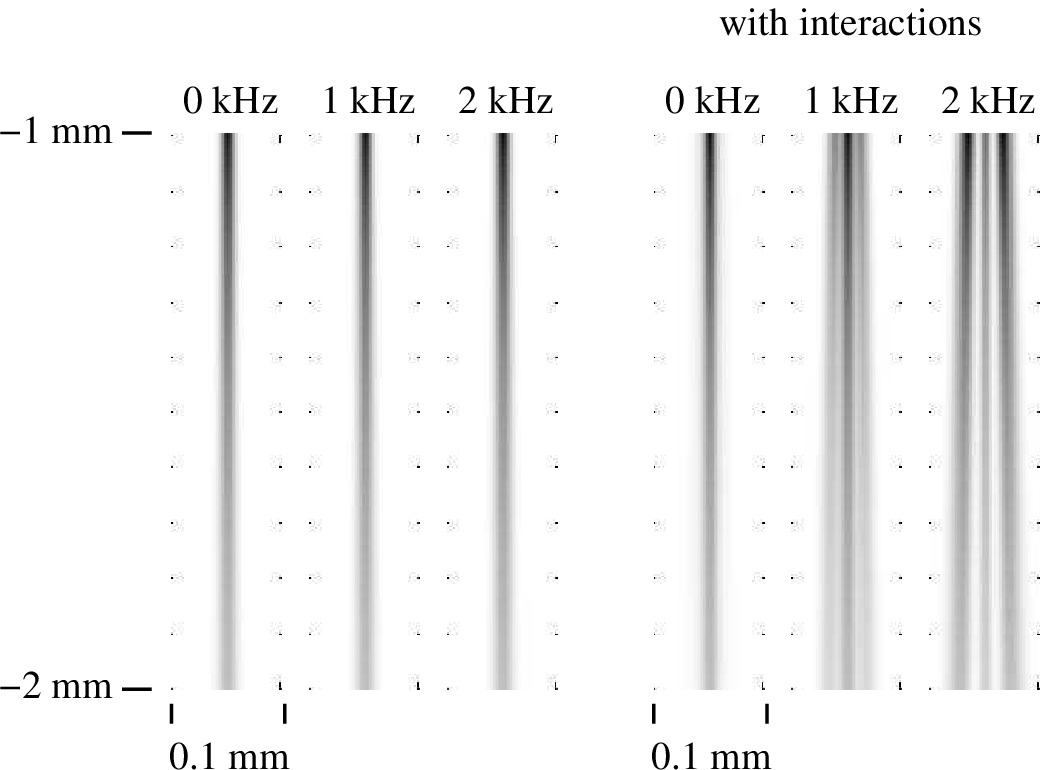}
\hfill
\includegraphics[width=0.52\textwidth]{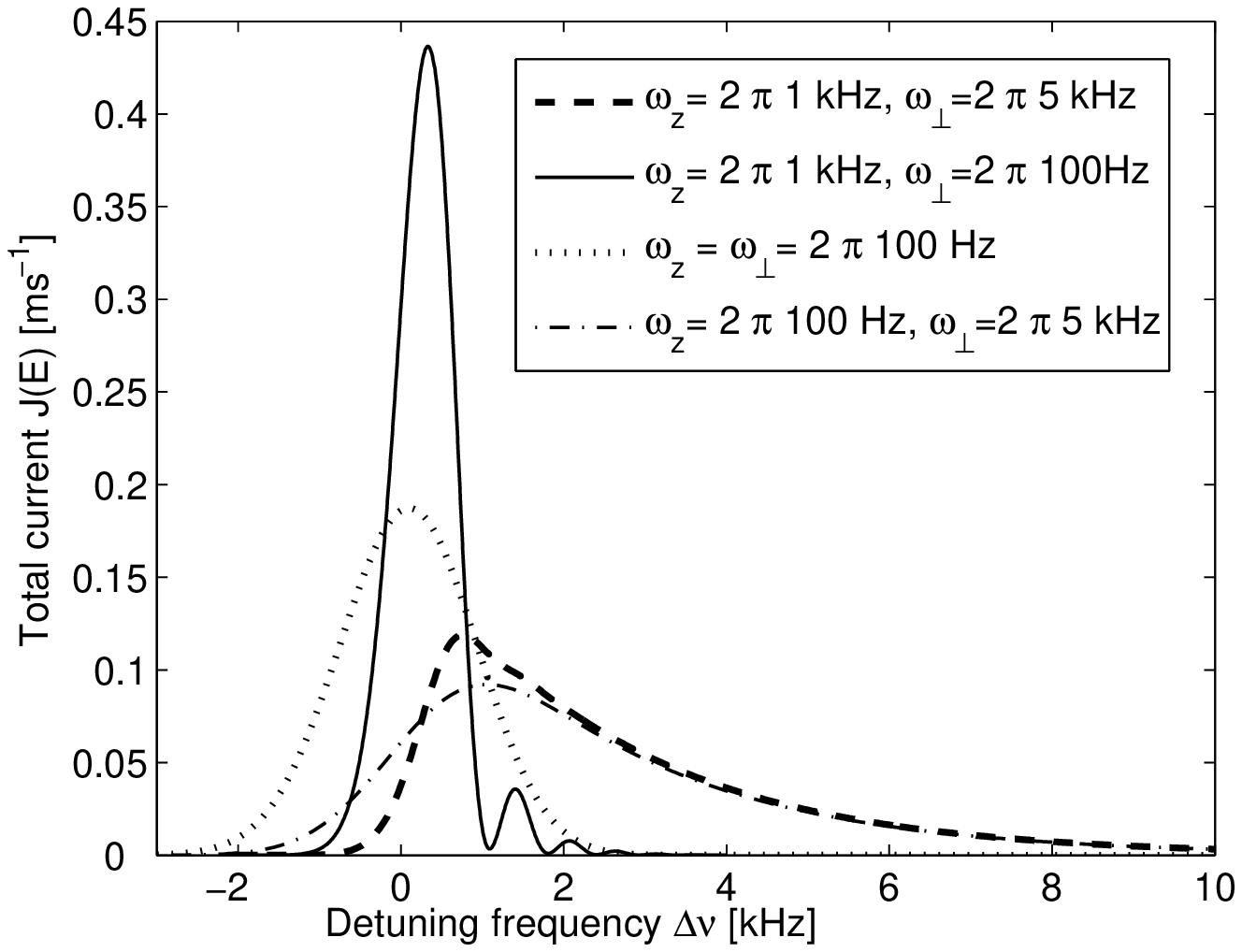}
\caption{
Left panels: Atom-laser beam profile from a Rb BEC of size $a=0.8\;\mu$m at different detuning energies $\Delta\nu=E/h$ (\ref{eq:PsiVirtual}). The first series shows the beam profile for non-interacting particles, whereas the next series includes interactions due to 500 atoms, which lead to a transverse substructure \cite{Kramer2006b}. Right panel: Anisotropic trapping frequencies cause a strong modulation
of the total particle current $J(E)$.
\label{fig:interaction}}
\end{figure}

For ideal, non-interacting atoms, the ultracold cloud populates the ground state of the nearly parabolic trapping potential, leading to a Gaussian density profile in the BEC.  For simplicity, we assume an isotropic distribution:
\begin{equation}
\label{eq:SourceGauss}
\sigma(\mathbf{r})=\hbar\Omega\psi_0(\mathbf{r})=\hbar\Omega N_0 \exp(-r^2/(2a^2)).
\end{equation}
Here, $\hbar\Omega$ denotes the strength of the transition-inducing oscillating RF field.  The parameter $a$ describes the width of the source (which is related to the field gradient in the trap), and $N_0=a^{-3/2}\pi^{-3/4}$ denotes the proper normalization from the condition
\begin{equation}
\int\rmd^3\mathbf{r}\,{|\psi_0(\mathbf{r})|}^2=1.
\end{equation}
To obtain expressions for the currents generated by a Gaussian source, we work in the time-dependent propagator representation (see Eq.~(\ref{eq:bkk9})).  The beam wave function $\psi(\mathbf r)$ then may be written
\begin{equation}
\label{eq:extend}
\psi(\mathbf{r})=-\rmi\Omega N_0\int_0^\infty \rmd t\,\rme^{\rmi E t/\hbar}\int \rmd^3\mathbf{r}'\,K_{\rm field}(\mathbf{r},t|\mathbf{r}',0)\,\rme^{-r'^2/(2a^2)}.
\end{equation}
It is possible to carry out the integration over the source volume.  With negligible corrections, outside the source the integral (\ref{eq:extend}) assumes the form
\begin{equation}
\label{eq:PsiVirtual}
\psi(\mathbf{r}) = \hbar\Omega {(2\sqrt{\pi}a)}^{3/2} \rme^{-m a^2 E/\hbar^2+m^2 F^2 a^6/(3\hbar^4)}
\,G(\mathbf{r},-\frac{m\mathbf{F}}{2\hbar^2}a^4;E),
\end{equation}
where $G(\mathbf{r},\mathbf r';E)$ denotes the energy Green function for uniformly accelerated particles (\ref{GreenField}).  This expression displays a remarkable feature of the beam wave function $\psi(\mathbf{r})$ originating from a Gaussian source: The extended source can be formally replaced by a virtual point source of the same energy, albeit at a location shifted by $\mathbf r' = -m\mathbf{F}a^4/(2\hbar^2)$ from the center of the Gaussian distribution (see Fig.~\ref{fig:virtual+beam}).  Expressions for the beam profile and currents are then conveniently found from the analogous expressions for a point source by performing the indicated shifts.

As an immediate, and somewhat surprising, consequence of the concept of a virtual source, the beam profile shows a sharp fringe pattern that results from the interference between the two virtual paths in Fig.~\ref{fig:virtual+beam}.  The number of fringes depends sensitively on the size $a$ of the source, as displayed in Fig.~\ref{fig:virtual+beam}.  In the limit of \emph{extended} Gaussian sources with $E < mF^2a^4/(2\hbar^2)$, the virtual source turns into a tunneling source (as discussed in greater detail in the following section), and the beam profile itself becomes Gaussian \cite{Bracher1998a}.  The spectrum of the total particle current $J(E)$ as a function of the detuning of the RF field $E = h\Delta\nu$ then may be written in the suggestive form
\begin{equation}
\label{eq:CurrentTotalSP}
J(E) \approx \frac{2\pi}{\hbar}\int\rmd^3\mathbf{r}\,{|\sigma(\mathbf{r})|}^2\,\delta(E+ Fz), 
\end{equation}
an expression that has a simple geometrical interpretation. For extended sources, the energy dependence of $J(E)$ reflects the source structure: By the resonance condition $E+Fz=0$, the total current probes the density ${|\psi_0(\mathbf{r})|}^2$ of the BEC on different slices across the source.  Finally, we note that the approximation (\ref{eq:CurrentTotalSP}) obeys the sum rule (\ref{eq:Multi1.9}) for the total current $J(E)$.

In an actual atomic BEC, the repulsive interactions between atoms lead to a broadening of the condensate. For most cases, the inclusion of the interactions via a mean-field approach is sufficient. The repulsive forces of the much denser BEC act on the outcoupled atom beam and lead to a further splitting of the beam profile, as shown in Fig.~\ref{fig:interaction}. Also the total current is modified by the interactions \cite{Kramer2006b}. Both effects have been observed experimentally. Non-isotropic trapping frequencies and currents from higher trapping modes allow to control the shape and rate of the atom laser \cite{Kramer2006b}.

\begin{figure}
\begin{center}
\includegraphics[width=0.38\textwidth]{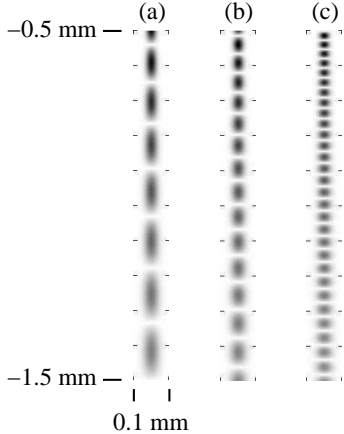}
\caption{
Beam profile for simultaneous output coupling with
two different radio frequencies. The outcoupling frequencies
$\Delta\nu_{1,2}=E/h$ are (a) $\pm 0.5$~kHz, (b) $\pm 1.0$~kHz, and (c)
$\pm 2.0$~kHz. The number of longitudinal interference  fringes is
proportional to the difference in the detuning frequencies.
Parameter: $a=0.8\;\mu$m, $F=m_{\text{Rb}}\,g$, with
$g=9.81$~m/s$^2$, and
$m_{\text{Rb}}=87$~u.
\label{fig:faucet}
}
\end{center}
\end{figure}

Fig.~\ref{fig:faucet} shows the ``dripping quantum faucet'', which is produced by superposition of two laser beams with slightly different energy that are outcoupled from the same BEC \cite{Bloch2000a}. It is not surprising to see that rotating BECs which sustain vortices are described in terms of rotating Gaussian sources with nodal structures \cite{Bracher2003a}.

\section{Ballistic tunneling: STM}

\begin{figure}[t]
\centerline{\includegraphics[width=0.6\textwidth]{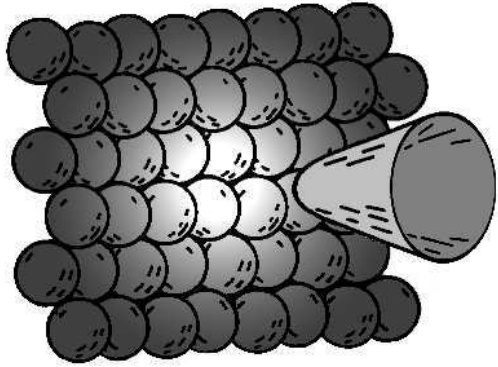}}
\caption{Electric field emission out of (or into) the sharp apex of a conducting wire. The bright spot symbolizes the Gaussian density profile along the central tunneling path between tip and surface of the sample.
\label{fig:kspot}}
\end{figure}

The quantum theory of scattering is not limited to asymptotic problems where particles are generated (and observed) far away from the scattering region. A prominent candidate for scattering at finite distances is the  Scanning Tunneling Microscope (STM). There, an electric current flows down a macroscopic wire that ends in a sharp tip. Its apex can be viewed as a source of electrons which leave the tip by tunneling due to the applied electric field between tip and sample surface. In some cases, the apex of the tip is ultra-sharp, consisting of a single atom. In an experiment the tip is slowly moved across the surface. In the constant current mode the tip is raised and lowered so as to keep the current constant. The raising and lowering process produces a computer-generated contour map of the surface \cite{Binnig1982a,Chen1993a}.  The method is capable of resolving individual atoms and works best with conducting materials. Electrons drawn from the apex of the tip (see Fig.~\ref{fig:kspot}) exhibit dynamically forbidden motion because the electron transfer between tip and surface occurs via field-driven tunneling, confining the current to a narrow filament with Gaussian profile that samples the surface.
\begin{figure}[t]
\begin{center}
\includegraphics[width=0.6\textwidth]{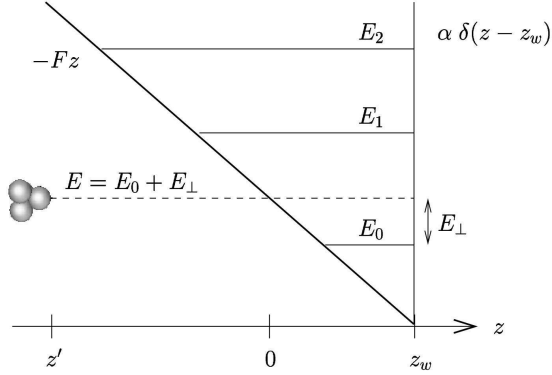}
\end{center}
\caption{Sketch of the bouncing ball problem. An electron exits the three-dimensional tunnel at $z=0$ with energy $E=0$. It then undergoes multiple reflections between the exit of the tunnel and the surface barrier at $z=z_w$ before travelling into the solid ($z>z_w$). Adatoms (not drawn in the figure) are necessary to observe the resonance-induced ripples (see Fig.~\protect{\ref{fig:fig2}}) with energy $E_\perp$.
\label{fig:res}}
\end{figure}
 
It is straightforward to model the apex of an STM tip as a source (or sink) of electrons. To be specific, 
we consider here a conducting sample surface that harbors a two-dimensional electron gas; in practice, the band of surface states on the densely packed, smooth Cu(111) surface has been exploited for this purpose \cite{Crommie1993a, Crommie1995a}.  The STM tip will emit a spreading surface electron wave that is scattered at adsorbed surface atoms (adatoms).  Since the electrons are slow, $s$--wave scattering prevails that can be modelled by a short-range potential. In this case, one has an analytic solution for the scattering problem which forms the basis for the calculation of the corrugation (surface roughness). As a result of the analytic approach we will show that scattering resonances play an essential role for resolving atoms and detecting electron surface states.
\begin{figure}
\begin{center}
\includegraphics[width=0.6\textwidth]{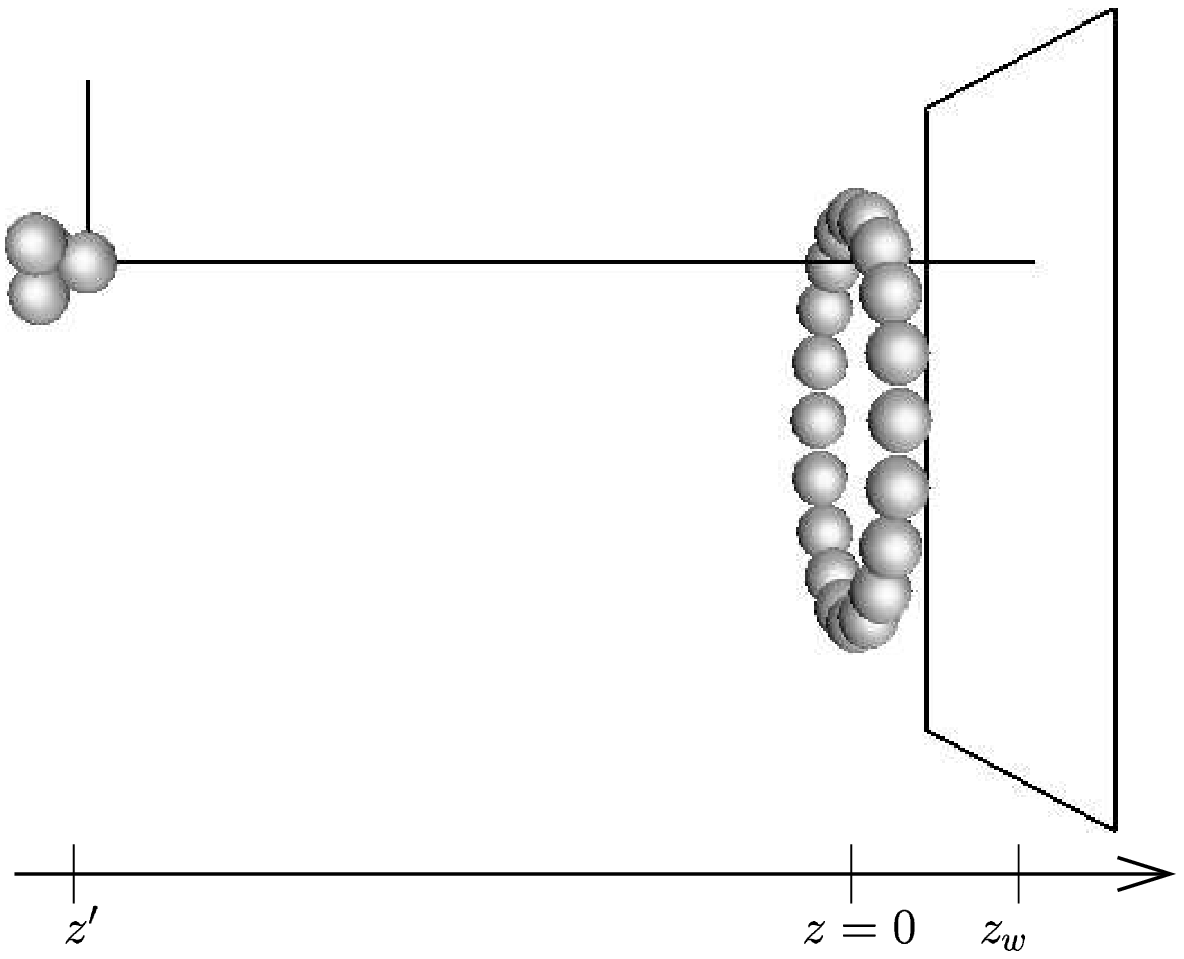}
\caption{Schematic plot of tip (at $z=z^\prime$), adsorbed atoms (at $z=0$), and a strongly reflecting, clean surface (at $z=z_w$).
\label{fig:scheme}}
\end{center}
\end{figure}
\begin{figure}
\begin{center}
\includegraphics[width=0.475\textwidth]{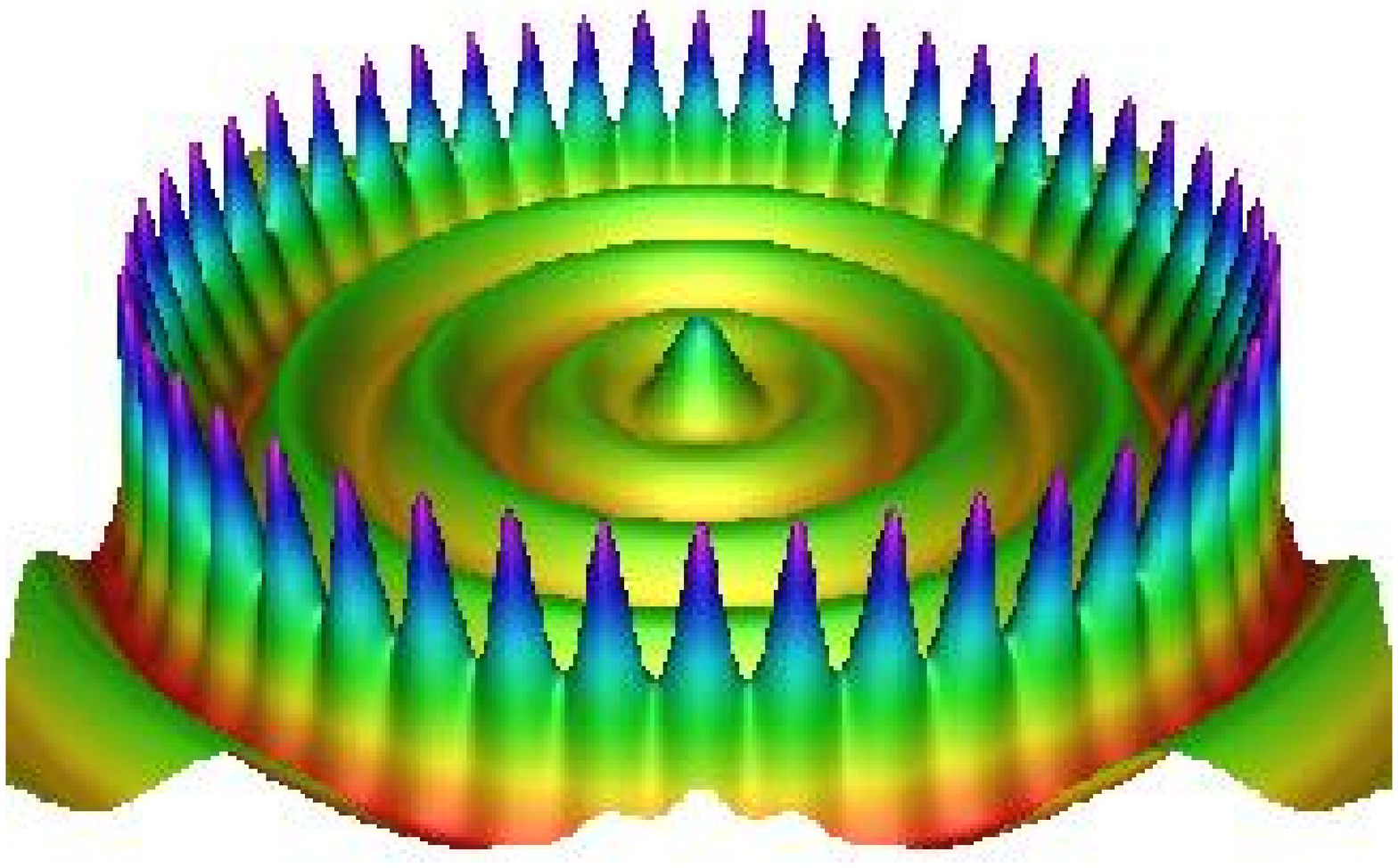}\hfil%
\includegraphics[width=0.475\textwidth]{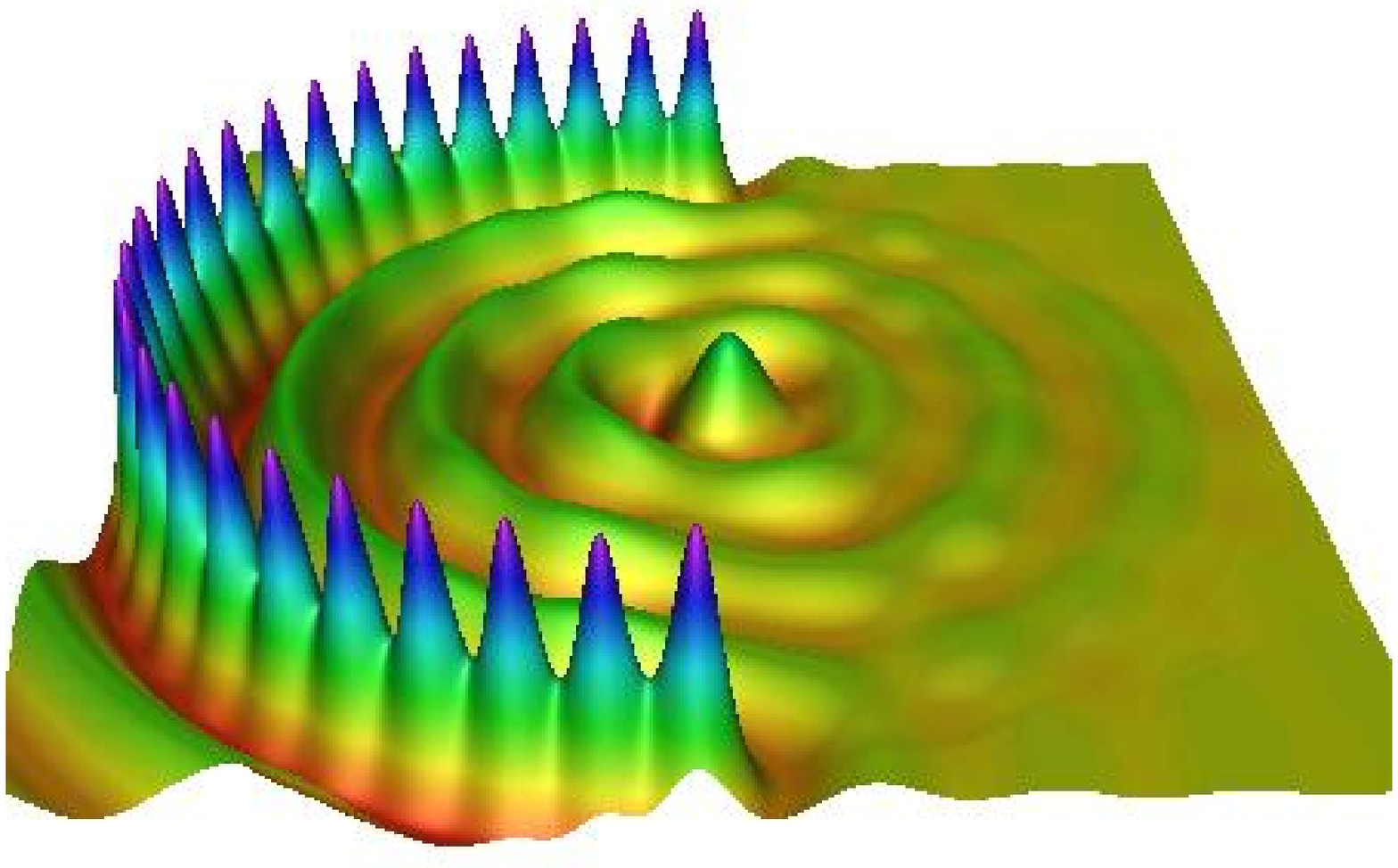}%
\end{center}
\caption{Interference of waves scattered by corral atoms: Model calculation \cite{Donner2005a} of a corrugation plot at constant electric current.
The structure of the circular ripples both inside and outside the corral can be related to the quantum bounce problem illustrated in Fig.~\protect{\ref{fig:res}} and discussed in the text. For comparison with experimental results obtained by Eigler's group \cite{Crommie1993a, Crommie1995a} we show a similar setup with a quantum corral consisting of 48 iron atoms on a circle with radius 71.3~\AA\ adsorbed on a Cu(111) surface (left panel). The corrugation of the adatoms is approximately 0.5~\AA\ and corresponds to a conductivity of $\sigma_0 = 2.7 \cdot 10^{-8}$ A/V. Note that there is no adsorbed atom in the center of the corral.
\label{fig:fig2}}
\end{figure}

The current flowing through the STM tip is proportional to the local density of surface states, and therefore the imaginary part of the Green function at the tip position, $\Im[G(\mathbf R,\mathbf R;E)]$ (\ref{eq:Multi1.11}).  A route that leads conveniently to the Green function in this problem consists of the following three steps:

\paragraph{Step 1. One-dimensional problem:}
We first calculate the one-dimensional Green function $G_1^\alpha(z,z';E)$ that corresponds to the model potential of Fig.~\ref{fig:res}. In this one-dimensional problem, the electron is allowed to tunnel in direction of the electric field ($z$--direction). The energy $E_{0}$ corresponds to the bound (and unoccupied) surface state of Cu(111). The energy $E$ of the tunneling electron is taken to lie in a band gap of the substrate (solid). As a result the electron faces a potential barrier at $z = z_{w}$ and bounces back and forth between tunnel exit ($z = 0$) and barrier. However, the electron can move freely with energy $ E_\perp = E - E_{0}$ in the surface plane orthogonal to  $z = z_{w}$. Because of inelastic scattering with phonons the electron will finally disappear in the solid. By assuming a point source at $z = z'$ the Green function that belongs to the problem of Fig.~\ref{fig:res} can be solved analytically in terms of Airy functions \cite{Donner2005a}.

\paragraph{Step 2. Three-dimensional background Green function:}
Since the STM is a three-dimensional device we must calculate the Green function in three dimensions. The uncertainty principle for momentum and position applied in the lateral ($x-y$) direction results in a tunneling spot (see Fig.~\ref{fig:kspot}) of finite width, and approximately Gaussian profile. The three-dimensional Green function $G_\text{sym}(\mathbf{r},\mathbf{r}', z)$ for a particle moving in the potential $\tilde{U}(\mathbf{r})=\tilde{U}(z)$ of Fig.~\ref{fig:res} is obtained from its one-dimensional counterpart by integrating $G_1^\alpha$ over all momenta $\hbar k_\perp$ vertical to the field direction (i.~e., parallel to the surface),
\begin{equation}
\label{eq:eq10}
G_\text{sym}(\mathbf{r}, \mathbf{r'}, E) = G_\text{sym}(z ,z',\,
\Delta\rho, E) = \frac{1}{2\pi}\int_0^\infty {\rm d}k \, k_\perp \, J_0(k_{\perp} \, \Delta \rho)\, G_1^\alpha \Bigr(z, z', E-\frac{\hbar^2k_\perp^{2}}{2 M}\Bigl),
\end{equation}
with $J_0(\cdots)$ being the usual cylindrical Bessel function of degree zero. The lateral distance between $\mathbf{r}$  and $\mathbf{r}'$ is given by $\Delta\rho^2 =|\rho - \rho'|^2 = (x-x')^2 + (y-y')^2$. Obviously $G_\text{sym}(\mathbf{r'}, \mathbf{r'}, E)$ is independent of the lateral position $(x',y')$ of the tip. Hence, for $z'=\text{const}$, $G_{\text{sym}}$ represents a constant background corrugation. The full solution for the Green function with the adatoms present, is then obtained from

\paragraph{Step 3. Dyson equation for $G(\mathbf{r}, \mathbf{r'}, E)$:}
We must now take into account the adsorbed atoms (see Fig.~\ref{fig:scheme}). Using the appropriate Dyson equation, we obtain an algebraic equation for the full Green function,
\begin{equation}
\label{eq:eq14}
G(\mathbf{r}, \mathbf{r}', E) = G_\text{sym}(\mathbf{r},
\mathbf{r}', E) + \sum_{j,k=1}^{n} G_\text{sym}(\mathbf{r},
\mathbf{r}_j, E) (\mathbf{T}(E))_{jk}
G_\text{sym}(\mathbf{r}_k,\mathbf{r}', E),
\end{equation}
where the sum runs over all adatoms, and the $T$--matrix describing the effects of the multiple scattering events between the adsorbed atoms can be expressed using the background Green function $G_\text{sym}(\mathbf{r}_j, \mathbf{r}_k, E)$. Details of the calculation of the Green function and the experimentally observable tunneling current $J(\mathbf R;E)$ (\ref{eq:Multi1.11}) can be found in Ref.~\cite{Donner2005a}.  A zero-temperature plot obtained from such a calculation (which typically takes a few minutes on a personal computer) is shown in Fig.~\ref{fig:fig2}.  We should point out that since $G_{\text sym}$ gives rise only to a uniform background current, the observed roughness of the surface is entirely contained in the $T$--matrix. 

\section{Electrons in Electric and Magnetic Fields:\\ The Quantum Hall Effect}
\label{sec:QHE}

\begin{figure}[t]
\begin{center}
\includegraphics[width=0.5\textwidth]{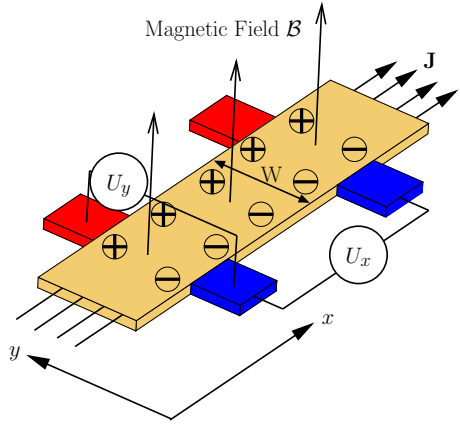}
\caption{Schematic view of a Hall bar. A current $\mathbf{J}_x$ is flowing through a two-dimensional electron gas (2DEG) in the $x$-$y$--plane, which is orientated perpendicular to an external magnetic field ${\cal B}$. The deflected electrons at the sample edges produce a Hall voltage $U_y$ over the sample width $W$, which is measured along with the longitudinal voltage drop $U_x$.
\label{fig:hallbar}}
\end{center}
\end{figure}

\begin{figure}
\begin{center}
\includegraphics[width=0.7\textwidth]{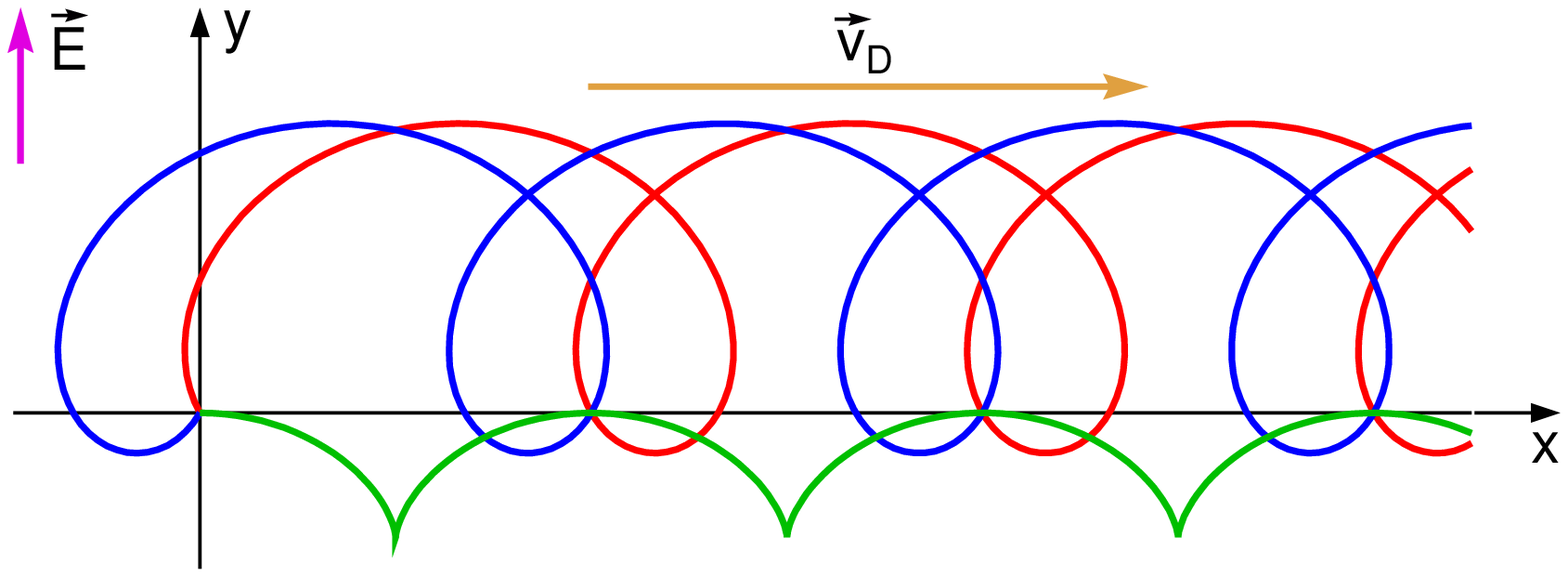}
\caption{Drift motion in crossed fields: Trajectories drift with velocity $v_D = {\cal E}/{\cal B}$ perpendicular to the fields. The motion of the electrons in two dimensions is a superposition of cyclotron motion with drift motion, resulting in \emph{trochoidal} (cycloidal) trajectories.
\label{fig:qhe1}}
\end{center}
\end{figure}

In this section we explore the strange and fascinating ways of electrons in electric and magnetic fields. Of particular importance here is the \emph{Hall configuration}, where the electrons are confined to an effectively two-dimensional conductor in the presence of orthogonal electric and magnetic fields.  The Hall geometry is displayed in Fig.~\ref{fig:hallbar}.
Fig.~\ref{fig:qhe1} shows some of the classical paths followed by the electrons in the conducting plane.  Notwithstanding the complicated pattern of motion, all trajectories share the same distinctive behavior, uniform \emph{drift motion} perpendicular to both fields with a characteristic velocity $v_D = {\cal E}/{\cal B}$.

Because of the universal drift motion, the current density in the Hall bar is simply proportional to the local density of states (LDOS) $n(E)$ in the material, which in turn is related to the Green function in the corresponding external potentials (see Sec.~\ref{sec:Multi4}).  Hence, from a mathematical point of view we are interested in finding the energy-dependent density of states for the moving electrons.  For a purely magnetic field, the two-dimensional LDOS has a spike-like structure \cite{Prange1987a,Grosso2000a}, formally written as a superposition of discrete $\delta$--distributions positioned at the Landau levels at $E=(2k+1)\hbar\omega_L$, where $\omega_L=e\mgf /(2m)$ denotes the Larmor frequency.
\begin{equation}
\label{eq:DOSB}
n_{\MGF}^{(2D)}(E)=\frac{e \mgf }{2\pi\hbar}\sum_{k=0}^\infty
\delta\left(E-\hbar\omega_L[2k+1]\right).
\end{equation}

The addition of an electric field leads to important changes in the density of states for a purely magnetic field. As we know from Eq.~(\ref{eq:Multi1.11}), the local density of states $n(E)$ is always linked to the imaginary part of the energy-dependent retarded Green function $G(\mathbf{r}=\mathbf{o},\mathbf{r}'=\mathbf{o};E)$,
\begin{equation}
\label{eq:LDOS}
n(E)=-\frac{1}{\pi}\Im\left[G(\mathbf{o},\mathbf{o};E)\right],
\end{equation}
which in turn can be expressed as the Laplace transform of the quantum propagator $K(\mathbf{o},t|\mathbf{o},0)$ (\ref{eq:bkk9}):
\begin{equation}
\label{eq:Green}
G(\mathbf{o},\mathbf{o};E) = \frac{1}{{\rm i}\hbar} \int_0^\infty {\rm d}t\,
{\rm e}^{{\rm i}E t/\hbar} K(\mathbf{o},t|\mathbf{o},0).
\end{equation}
For $\mathbf{r}=0$, the two-dimensional quantum propagator for the Hamiltonian of a (spinless) electron in crossed fields,
\begin{equation}
\label{eq:HEB}
H_{\ELF\times\MGF}^{(2D)} =  \frac{\mbfp_x^2+\mbfp_y^2}{2m} +\frac{1}{2}m\omega_L^2\left(x^2+y^2\right) - \mathbf r_\perp\cdot\mathbf F_\perp -\mbfp_y x \omega_L +\mbfp_x y \omega_L,
\end{equation}
where $\mathbf F_\perp = -e \mathbf{\elf}$ denotes the electric force in the $x$-$y$--plane reads \cite{Grosche1998a},
\begin{equation}
\label{eq:JCrossedInt2D}
K_{\ELF\times\MGF}^{(2D)}(\mbfo,t|\mbfo,0)=
-\frac{\rmi m\omega_L}{2\pi\hbar\sin(\omega_L t)}
\exp \left\{ \frac{\rmi F_\perp^2 t}{8m\hbar\omega_L^2}
\left[ \omega_Lt \cot\left(\omega_{L}t\right) - 1 \right]
\right\}.
\end{equation}
To account for the effects of the electron spin, we note that its interaction with the magnetic field merely adds a spatially constant term to the Hamiltonian, and thus shifts the effective energy of the two spin populations by a fixed amount $\pm \frac{1}{2}g \mu_B\mgf = \pm \frac{1}{2}g\hbar\omega_L$ \cite{Prange1987a}. The spin-dependent densities of states become
\begin{equation}
\label{eq:spin}
n_{\uparrow,\downarrow}(E) =n\left(E \pm \frac{1}{2}g\hbar\omega_L\right)
\end{equation}
and the total LDOS including spin can be mapped back to the LDOS without spin: $n_{\uparrow\downarrow}(E)=n_\uparrow(E)+n_\downarrow(E)$.  Thus, it suffices to evaluate $n(E)$.  
For the discussion of the density of states in the Hall configuration it is useful to replace the trigonometric functions in the propagator (\ref{eq:JCrossedInt2D}) by a sum. This can be done using the identity
\begin{equation}
\frac{\exp[-\alpha\coth(z)]}{\sinh(z)} = 2\rme^{-\alpha} \sum_{k=0}^\infty
\La_k^{(0)}(2\alpha)\rme^{-2z(k+1/2)},
\end{equation}
which follows directly from the generating function of the Laguerre polynomials $\La_k^{(0)}(z)$ \cite{Abramowitz1965a}.  The Laplace transform (\ref{eq:Green}) then can be performed analytically, and we finally find for the density of states
\begin{equation}
\label{eq:DOSEB}
n_{\ELF\times\MGF}(E) = \frac1{2\pi^{3/2}l^2\Gamma} \sum_{k=0}^\infty
\frac{1}{2^k k!} \, \rme^{-E_k^2/\Gamma^2} {\left[\He_k\left(E_k/\Gamma\right)\right]}^2 ,
\end{equation}
where the level width parameter
\begin{equation}
\Gamma = F_\perp l
\end{equation}
is related to the magnetic length $l = \sqrt{\hbar/(e\mgf)}$. $\He_k(z)$ denotes the $k$th Hermite polynomial \cite{Abramowitz1965a} and $E_k$ is the effective energy shift for the $k$th Landau level
\begin{equation}
E_k=E-\Gamma^2/(4\hbar\omega_L)-(2k+1)\hbar\omega_L.
\end{equation}
Interestingly, the density of states can again be interpreted as a sum over Landau levels.  However, they now appear spread in energy, with a distribution that is isomorphic to the probability density of the corresponding eigenstate of a one-dimensional harmonic oscillator
\begin{equation}
\label{eq:QHO}
{\left|u_k(\xi)\right|}^2 = \frac1{2^k k!\,\sqrt\pi} \,
\rme^{-\xi^2}
{\left[ \He_k\left(\xi\right)\right]}^2 .
\end{equation}
The \emph{total} contribution of the $k$th Landau level integrated over energy-space is readily available from the normalization of the oscillator eigenstates:
\begin{equation}
\label{eq:EBQ}
\int_{-\infty}^\infty\rmd E\;
n_{k,\ELF\times\MGF}(E)=
\frac{e\mgf }{2\pi\hbar}
\int_{-\infty}^\infty\rmd \xi\;{\left|u_k(\xi)\right|}^2 = \frac{e\mgf }{2\pi\hbar}.
\end{equation}
This result reflects the quantization of each Landau level in a purely magnetic field (\ref{eq:DOSB}).

\begin{figure}[t]
\begin{center}
\includegraphics[width=0.41\textwidth]{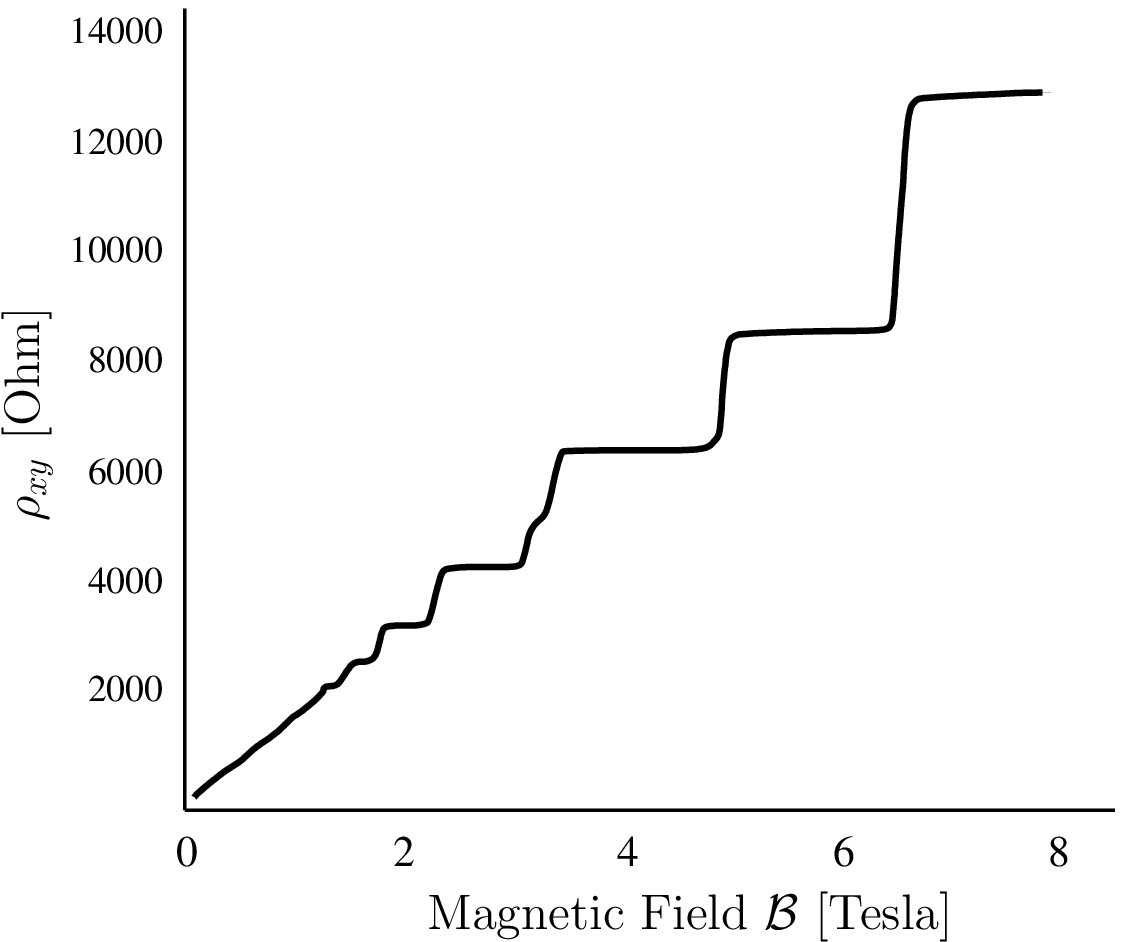}%
\hfil%
\includegraphics[width=0.55\textwidth]{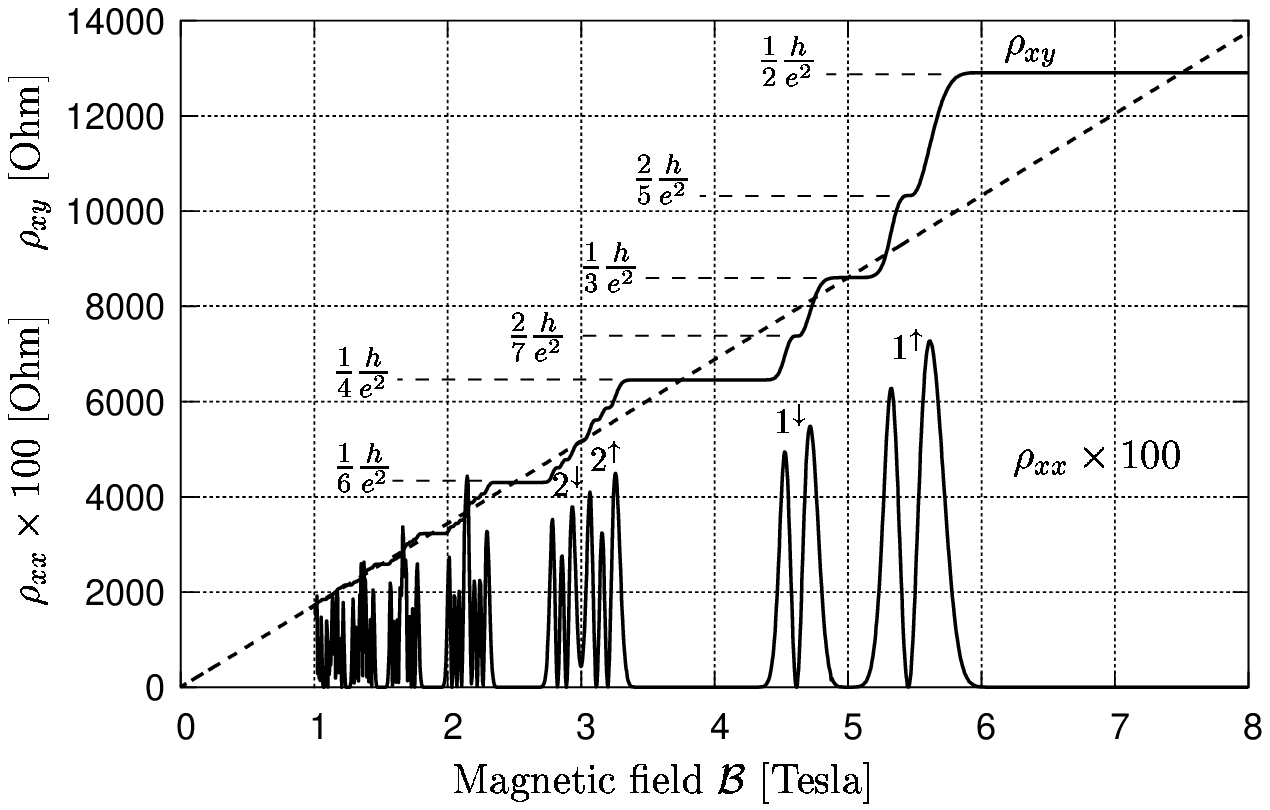}%
\hfill 
\caption{The quantum Hall effect.  Left panel: Schematic sketch of the experimental results adapted from \cite{Paalanen1982a} for $\rho_{xy}$ in GaAs/AlGaAs heterojunctions at $T=50$~mK. The theoretical simulations \cite{Kramer2003a} (right panel) are based on the density of states (\ref{eq:DOSEB}). The dotted straight line is the classical prediction.
\label{fig:ourqhe}}
\end{center}
\end{figure}

At low temperatures, only the \emph{occupied states} with energies smaller than the Fermi energy $E_F$ of the system will contribute to the Hall current.  Thus, we expect that the current density will be proportional to the integrated density of states $N(E_F)$ with energies $E < E_F$,
\begin{equation}
\label{eq:IDOS}  
N(E_F)=\int_{-\infty}^{E_F}\rmd E\;n(E).
\end{equation}
Within the Fermi gas model for a dilute gas of electrons in two dimensions it is straightforward to obtain 
resistivity plots which bear remarkable resemblance to actual quantum Hall data \cite{Kramer2003a}.  An example is shown in Fig.~\ref{fig:ourqhe}. 

We finally mention the implications of our Green function model of the integer quantum Hall effect for the observed breakdown of the quantized conductivity at higher electric fields \cite{Klitzing1980a}. Kawaji \cite{Kawaji1993a,Kawaji1996a,Shimada1998a} studied the width of the quantized resistivity ``plateaus'' as a function of the electric current and thus the Hall field in the sample. He finds a characteristic power law for the shrinking of the plateaus, which can be expressed in terms of a critical electric field:
\begin{equation}\label{eq:crit}
\Elf_{\rm crit}\propto B^{3/2}. 
\end{equation}
The non-perturbative inclusion of the electric field via the Green function formalism as presented here leads automatically to this power law (see \cite{Kramer2003a,Kramer2003b,Kramer2003d,Kramer2005c}) 

\section{The semiclassical method}
\label{sec;Semi}

It is usually rewarding to analyze quantum problems with the tools of classical physics.  First, the semiclassical solution facilitates in many cases the understanding of the quantum solution properties. Second, semiclassics leads to approximate solutions (``WKB solutions") that are usually numerically much less expensive than \emph{ab initio} quantum mechanical calculations.  In context of the propagator methods discussed here the semiclassical method can be described by the following steps: 
\begin{enumerate}
\item
Find \emph{number} of paths $N$ joining source $\mathbf r'$ and destination $\mathbf r$. This is an effective method for finding the caustics (turning surfaces).
\item
Find all trajectories $\mathbf r_k(t)$ leading from $\mathbf r'$ to $\mathbf r$.
\item
Establish \emph{classical weight} $\rho_k(\mathbf r)$, i.~e., the local density of trajectories, for each path.
\item
Determine its \emph{semiclassical phase} $\Phi_k(\mathbf r)$ from the reduced action along the path. 
\item
Create \emph{semiclassical approximation} to the Green function $G_{\rm sc}(\mathbf r, \mathbf r'; E)$.
\item
Use \emph{uniform approximations} to correct the divergence of the WKB solution near the caustics. 
\end{enumerate}
Therefore, the semiclassical method requires knowledge of the classical trajectories $\mathbf r_k(t)$, their local density $\rho_k(\mathbf r)$, and the corresponding action fields $W_k(\mathbf r,\mathbf r'; E)$.  A semiclassical treatment of the photodetachment problem (Sec.~\ref{sec:Multi4}) can be found e.~g.\ in Ref.~\cite{Bracher1998a}.  Recently, the technique has been applied to the more complicated dynamics of electrons in parallel electric and magnetic fields \cite{Bracher2006a,Bracher2006b}.  Here, we concentrate on the related problem of semiclassical motion in the Hall configuration (Sec.~\ref{sec:QHE}).

\begin{figure}[t]
\begin{center}
\includegraphics[width=0.7\textwidth]{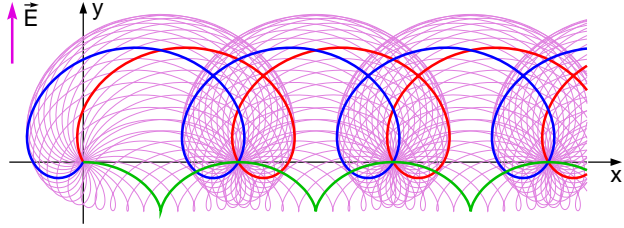}
\caption{Caustics and the trajectory field (cf.\ Fig.~\ref{fig:qhe1}): The trajectory field traces out the caustics. Electrons pass through focal points. The number of paths connecting the source with a given point increases with the magnetic field.
\label{fig:crossed}}
\end{center}
\end{figure}

\paragraph{Classical orbits for the Hall effect:}
The propagation of electrons in crossed electric and magnetic fields has received much attention in classical physics. Indeed, the corresponding trajectory field of electrons emitted by a point source located at the origin (for convenience), plotted in Fig.~\ref{fig:crossed}, looks interesting by itself.  Again assuming the Hall geometry displayed in Fig.~\ref{fig:hallbar}, the motion of the charges is governed by the Hamiltonian (\ref{eq:HEB}), and we find the family of orbits:
\begin{equation}
\label{eq:traj}
\mathbf r(t) = \begin{pmatrix} v_D t \\ 0 \end{pmatrix} + 
\frac 1{2\omega_L}
\begin{pmatrix} -v_{0y} & v_{0x} - v_D \\ v_{0x} - v_D & v_{0y} \end{pmatrix}
\cdot
\begin{pmatrix} \cos(2\omega_L t) \\ \sin(2\omega_L t) \end{pmatrix}
+ \frac 1{2\omega_L} \begin{pmatrix} v_{0y} \\ v_D - v_{0x} \end{pmatrix}.
\end{equation}
Here, the initial velocity vector depends on the emission angle $\theta$ and is given by
\begin{equation}
\dot{\mathbf{r}}(t=0) = \mathbf v_0 = \begin{pmatrix} v_{0x} \\ v_{0y} \end{pmatrix}
= \begin{pmatrix} v_0 \cos\theta \\ v_0 \sin\theta \end{pmatrix}.
\end{equation}
It must fulfil $E = \frac12 mv_0^2$.  Eq.~(\ref{eq:traj}) is conveniently interpreted as a sum of three terms:  On average, only the first term contributes to the transport of the electron.  The corresponding drift velocity (averaged over one cyclotron period $T=\pi/\omega_L$) reads
\begin{equation}
\label{eq:DriftVelocity}
\mathbf{v}_D = \frac{1}{T}\int_{t}^{t+T}{\rm d}t'\,
\dot{\mathbf{r}}(t')=(\ELF\times\MGF)/\mgf^2.
\end{equation}
In a ``drift'' reference frame that is moving with this velocity the otherwise trochoidal orbit becomes a circle with angle-dependent radius $R(\theta)$ whose center is shifted from the origin by a constant displacement $\mathbf r_c$
\begin{equation}
\label{eq:CyclotronRadius}
R(\theta)^2 = \frac{v_0^2 - 2v_0v_D\cos\theta + v_D^2}{4\omega_L^2} \;, \qquad
\mathbf r_c = \frac 1{2\omega_L}\begin{pmatrix} v_0\sin\theta \\ v_D - v_0\cos\theta \end{pmatrix}.
\end{equation}
Some sample orbits are plotted in Fig.~\ref{fig:qhe1}.  Variation of the angle $\theta$ yields the trajectory field displayed in Fig.~\ref{fig:crossed}.

\paragraph{The Hall conductivity tensor:}
Noting that the velocity in turn is related to the classical current density $\mathbf{j}$
\begin{equation}
\label{eq:ClassicalHall}
\mathbf{j}=N e \mathbf{v}_D,
\end{equation}
where $N$ denotes the electron density, one can extract the resistivity tensor $\mbfrho$ (or its inverse, the conductivity tensor $\mbfsigma$) from Ohm's law in the two-dimensional $(x,y)$--plane:
\begin{equation}
\label{eq:ConductivityTensor}
\mathbf{j}=\mbfrho^{-1}\cdot\ELF
\quad\Rightarrow\quad
\mbfsigma = \mbfrho^{-1} =
\frac{N e}{\mgf}
\begin{pmatrix} 0 & -1 \\ 1 & 0 \end{pmatrix}
\end{equation}
This remarkable equation predicts a finite conductivity even in the absence of scattering, which is usually invoked in theories of conduction in order to guarantee a finite carrier velocity. We note that the drift velocity is independent of the kinetic energy of the electrons.

\paragraph{Closed orbits in the classical picture:}
Eq.~(\ref{eq:ClassicalHall}) shows that the current in the Hall conductor is determined by the density of states $N$.  According to Sec.~\ref{sec:Multi}, the local density of states (LDOS) $n(E)$ within a narrow energy interval is related to the Green function $G(\mathbf o, \mathbf o;E)$ (\ref{eq:LDOS}), which, from a semiclassical perspective, is governed by those trajectories that return to the source ($\mathbf r = \mathbf r' = \mathbf o$).  This is the basic motivation of the closed orbit theory \cite{Berry1972a,Peters1993a} for source processes.

\begin{figure}[t]
\centerline{\includegraphics[width=0.7\textwidth]{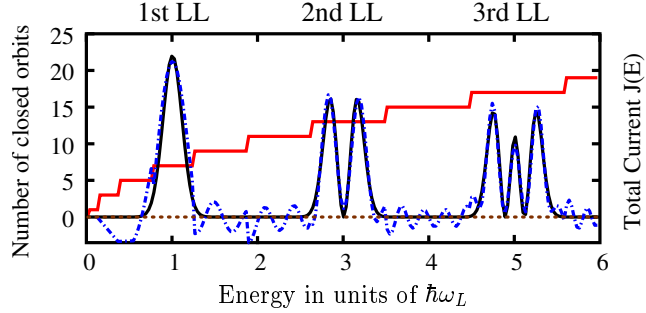}}
\caption{
Density of states in crossed electric and magnetic fields. The curves show the quantum (solid) and semiclassical result (dashed line). The staircase structure denotes the count of closed orbits. Crossed electric field $\Elf=4000$~V/m and magnetic field $\mgf=5$~T.
\label{fig:compare}}
\end{figure}

These trajectories, which lead from the origin back to the origin, are best found by using the classical action.  Here, it is convenient to start with the \emph{time-dependent} action functional $S_{\rm cl}(\mathbf r,t|\mathbf r',0)$.  In crossed electric and magnetic fields, this classical action is uniquely given by
\begin{equation}
\label{eq:SclExB}
S_{\rm cl}(\mathbf{o}, t|\mathbf{o},0)=
- \frac{m}{2} v_D^2 t
+ \frac{m\omega_L}2 \cot(\omega_L t) v_D^2 t^2.
\end{equation}
This expression describes the single closed orbit returning to the source in a predetermined time of flight $t$.  However, we are rather interested in the energy $E$ of the electron,
\begin{equation}
\label{eq:SaddlePoints}
E(t) = -\frac{\partial S_{\rm cl}(\mathbf{o},t|\mathbf{o},0)}{\partial t}.
\end{equation}
For fixed emission energy $E$, this is an implicit equation for the time of flight $t$, and generally several solutions $t_k$, pertaining to distinct classical trajectories $\mathbf r_k(t)$, exist.  (To find their initial velocities $\mathbf v_0$, it is sufficient to set $\mathbf r= \mathbf o$ and $t = t_k$ in the equation of motion (\ref{eq:traj}), and solve the ensuing linear equation system for $v_{0x}$ and $v_{0y}$.)  The reduced action for each contributing path then follows from the Legendre transform:
\begin{equation}
\label{eq:Legendre}
W_k(\mathbf o,\mathbf o;E) = S_{\rm cl}(\mathbf{o},t_k|\mathbf{o},0) + Et_k.
\end{equation}
As shown in Fig.~\ref{fig:compare}, the number of closed orbits increases with the magnetic field strength.

\paragraph{Density of states and propagator:}
The classical action (\ref{eq:SclExB}) is an important ingredient of the quantum-mechanical time-evolution operator. Using Eqs.~(\ref{eq:LDOS}) and (\ref{eq:Green}), it is possible to relate the local density of states with the propagator via \cite{Kramer2003a,Berry1972a,Gutzwiller1990a}
\begin{equation}
\label{eq:DOSIntegralExB}
n_{\ELF\times\MGF}^{(2D)}(E) = \frac{1}{2\pi\;\hbar} \int_{-\infty}^\infty \rmd T\, \rme^{\rmi ET/\hbar}\, K(\mathbf o, T|\mathbf o,0),
\end{equation}
where the time-dependent propagator (\ref{eq:JCrossedInt2D}) is given by
\begin{equation}
\label{eq:SCProp}
K(\mathbf{o}, t|\mathbf{o},0) = \frac{m\omega_L}{2\pi\rmi\;\hbar\sin(\omega_L t)}
\exp \left\{ \frac\rmi\hbar \, S_{{\rm cl}}(\mathbf{o}, t|\mathbf{o},0)\right\}.
\end{equation}
As we have shown before, this expression can be evaluated analytically in terms of harmonic oscillator eigenstates in energy space (\ref{eq:DOSEB}).
\begin{figure}[t]
\centerline{\includegraphics[width=0.7\textwidth]{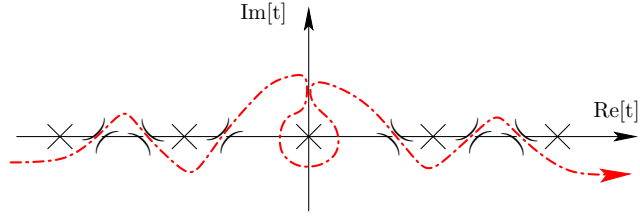}}
\caption{
Principal structure of the classical action in the complex time plane for $N=3$ saddle points (closed orbits). The dashed line denotes the integration path. Singularities are denoted by $\times$ and saddle points by $)($. Note that the singularity at the origin arises from the prefactor in the propagator and not from the classical action at $t=0$.
\label{fig:integrationpath}}
\end{figure}

\paragraph{Quantum result versus semiclassical approach:}
An asymptotic evaluation of the integral~(\ref{eq:DOSIntegralExB}) provides the link between closed orbits and the density of states. The original path of integration follows the real time-axis. Analytic continuation of the propagator makes it possible to deform this path of integration to the one sketched in Fig.~\ref{fig:integrationpath}. This path passes through saddle points of the exponent (denoted by $)($ in the figure) using the paths of steepest descent. The singularities in the integrand at times $T=k\pi/\omega_L$ (denoted by $\times$) are avoided. The only contribution of a singularity comes from $t=0$, which may be evaluated by the residue theorem:
\begin{equation}
I_{{\rm origin}}
=\frac{1}{2\pi\;\hbar}\oint\rmd t\, 
\rme^{\rmi E t/\hbar}\, K(\mathbf o, t|\mathbf o,0)
=\frac{m}{2\pi\;\hbar^2}.
\end{equation}
Comparison of Eqs.~(\ref{eq:DOSIntegralExB}) and (\ref{eq:SCProp}) with (\ref{eq:SaddlePoints}) shows that the saddle points of the integrand coincide with the classical times of flight for the various closed orbits.  Adding their contributions yields the semiclassical result:
\begin{equation}
n_{sc,\ELF\times\MGF}^{(2D)} = I_{{\rm origin}} + 
2\,{\rm Re}\left[ \frac{m\omega_L}{4\pi^2\rmi\;\hbar^2} \sum_{k=1}^N
\frac{\rme^{\rmi W_k(\mathbf o,\mathbf o;E)/\hbar + \rmi\pi \sgn[\ddot{S}_{cl}(\mathbf{o},t_k|\mathbf{o},0)]/4}}{\sin(\omega_L t_k)\,\sqrt{|\ddot{S}_{cl}(\mathbf{o},t_k|\mathbf{o},0)|/(2\pi\;\hbar)}} \right]
\end{equation}
Fig.~\ref{fig:compare} compares the semiclassical and quantum results. Despite their very different origins (sum over Landau levels vs.\ interfering classical trajectories), they are in striking agreement.  In the ``plateau regions'' of the conductivity, destructive interference between the properly weighted classical trajectories strongly suppresses the LDOS, leading to a quantization into separated levels with a substructure. Note that the number of trajectories does not change in each Landau level. It is the relative phase that modulates the LDOS.

\section{Acknowledgements.}

This work benefited greatly from discussions with W.~Becker, M.~Betz, B.~Donner, E.~Heller, M.~Moshinsky and M.~O.~Scully. Partial financial support from the Deutsche Forschungsgemeinschaft is gratefully acknowledged.

\end{document}